\documentclass[aps, prd, superscriptaddress, preprintnumbers, twocolumn, floatfix, nofootinbib]{revtex4-1}

\usepackage{amsfonts}
\usepackage{amsmath}
\usepackage{amssymb}
\usepackage{bm}
\usepackage{dcolumn}
\usepackage{epsfig}
\usepackage{graphicx}
\usepackage{graphics}
\usepackage[latin1]{inputenc}
\usepackage{latexsym}
\usepackage{rotating}
\usepackage{hyperref}
\usepackage{color}
\usepackage{lipsum}
\usepackage{hyperref}
\usepackage[sort&compress]{natbib}
\usepackage{float}

\newcommand\be{\begin{equation}}
\newcommand\ba{\begin{eqnarray}}
\newcommand\ee{\end{equation}}
\newcommand\ea{\end{eqnarray}}

\newcommand{\dd}{\mathrm d}

\newcommand{\As}{\mathcal{A}_{_\mathrm{S}}}
\newcommand{\ns}{n_\mathrm{s}}

\begin{document}

\title{Modeling the large-scale power deficit \\
with smooth and discontinuous primordial spectra}

\author{Sandro D.~P.~Vitenti}
\email{sandro.vitenti@uclouvain.be}
\affiliation{Instituto de F\'\i{}sica - Universidade de Bras\'\i{}lia - UnB
Campus Universit\'ario Darcy Ribeiro - Asa Norte
Sala BT 297 - ICC-Centro
70919-970 Bras\'\i{}lia, Brazil.}
\affiliation{Centre for Cosmology, Particle Physics and Phenomenology,
Institute of Mathematics and Physics, Louvain University, 2 Chemin
du Cyclotron, 1348 Louvain-la-Neuve, Belgium.}

\author{Patrick Peter}
\email{peter@iap.fr}
\affiliation{Institut d'Astrophysique de Paris and Institut Lagrange de Paris\\
CNRS (UMR 7095) and Sorbonne Universit\'e,
98 bis boulevard  Arago, 75014 Paris, France.}
\affiliation{Department of Applied Mathematics and Theoretical Physics,
Centre for Mathematical Sciences, University of Cambridge, Wilberforce
Road, Cambridge CB3 0WA, United Kingdom.}

\author{Antony Valentini}
\email{antonyv@clemson.edu}
\affiliation{Augustus College, 14 Augustus Road, London SW19 6LN, United
Kingdom.}
\affiliation{Department of Physics and Astronomy, Clemson University, \\
Kinard Laboratory, Clemson, South Carolina 29634-0978, USA.}
\date{\today}

\begin{abstract}
We study primordial power spectra with a large-scale power deficit and
their effect on the standard $\Lambda$CDM cosmology. The standard
power-law spectrum is subject to long-wavelength modifications described
by some new parameters, resulting in corrections to the anisotropies in
the cosmic microwave background. The new parameters are fitted to
different datasets: Planck 2015 data for temperature and for both
temperature and polarization, the low-redshift determination of $H_0$,
and distances derived from baryonic acoustic oscillations. We discuss
the statistical significance of the modified spectra, from both
frequentist and Bayesian perspectives. Our analysis suggests motivations
for considering models that break scalar-tensor consistency, or models
with negligible power in the far super-Hubble limit. We present what
appears to be substantial evidence, according to the Jeffreys' scale,
for a new length scale around 2200~Mpc ($k^{-1} \sim 350$~Mpc) above
which the primordial (scalar) power spectrum is sharply reduced by about
20\%.
\end{abstract}


\maketitle

\section{Introduction}

Cosmological data is now accumulating at such a rate that the phrase
`precision cosmology' is often used to describe our current
understanding of the primordial universe \cite{Peter:2013avv}. As is
well known, the largest scales show features whose significance is
unclear and which remain controversial \cite{Schwarz:2015cma}. In
particular, the existence or otherwise of a large-scale power deficit
remains an open question. A natural way to address this, explored in the
present paper, is to postulate that the primordial power spectrum is
suppressed at large scales by some as-yet-unknown physical mechanism
\cite{Chluba:2015bqa}. Assuming a phenomenological parametrization of
the modification, which we shall from now on refer to as a deficit
function, we may then analyze the available data and evaluate the
statistical significance of any proposed modified spectrum.

A previous similar analysis of the Planck data from the cosmic microwave
background (CMB) was carried out by the Planck team
\cite{PlanckCollaboration2014, Ade:2015lrj}, specifically for their
temperature and polarization data. They considered, among other
possibilities, two modifications of the primordial power spectrum with
suppressed power on large scales modeled by two extra parameters. They
concluded that ``{\sl neither of these two models with two extra
parameters is preferred over the base $\Lambda$CDM model}''. In a more
recent analysis~\cite{2018arXiv180706211P}, other models (and features)
based on inflationary scenarios were also considered, for which they
again found no supporting evidence. Even so, observations and
statistical tests of this deficit have a much longer history. The first
hint of a large-scale power deficit was already present in the first
detection of the CMB anisotropies by the COBE DMR experiment (see the
four-year results summary in Ref. ~\cite{1996ApJ...464L...1B}). This
feature was later observed by WMAP~\cite{2003ApJS..148....1B}, and in
Ref.~\cite{2003ApJS..148..175S} they introduced the statistic $S_{1/2}$
to measure the lack of power at large angular scales ($>60{}^{\circ}$)
for the angular two-point correlation function and obtained a
moderate-to-strong significance for the low power (only~$0.15\%$ of the
simulations had the same low power). Following the WMAP result, several
different approaches were proposed to model and/or to test for low power
at large scales (see for example~\cite{2003JCAP...07..002C,
2003MNRAS.342L..72B, 2003MNRAS.343L..95E, 2003MNRAS.346L..26E,
2003PhLB..570..151K, 2004PhRvD..69f3516D, 2004PhRvD..69h3515M}). These
included spatially curved models \cite{Park:2018fxx}, modified
inflation, and purely phenomenological modifications of the primordial
power spectrum (PPS), among others. Besides the low power at large
angular scales, other so-called anomalies have also been considered (see
for example Ref. ~\cite{Muir:2018hjv} for a summary of these anomalies
and tests of their statistical significance). While statistical analysis
of ``anomalies'' can shed light on their significance, these \textit{a
posteriori} methods tend to overestimate the significance\footnote{The
correct significance could be obtained if the look-elsewhere effect were
taken into account. In most cases, however, it is not clear how to
compute this effect.}. A statistical test including the whole fit
provides a clearer picture of the significance of the proposed anomaly
(see for example Refs. ~\cite{2003JCAP...07..002C,
2011ApJS..192...17B}).

The purpose of this paper is to revisit the Planck team's conclusions by
considering a wider set of possible deficit functions together with a
wider set of cosmological data -- specifically including the
low-redshift determination of $H_0$ \cite{Riess2016} (leading to $H_0 =
73.24 \pm 1.74\, \hbox{km} \cdot \hbox{s}^{-1}\cdot\hbox{Mpc}^{-1}$
which is in tension\footnote{In this work, we find that correlations
between the $H_0$/Planck tension and the large-scale power deficit
appear only when temperature fluctuations alone are taken into account
and thus do not seem statistically relevant. See, however,
Ref.~\cite{2017PhRvD..96h3526O} for further details about the
relationship between cosmological parameters and anomalies in the CMB
data.} with the Planck result $H^\mathrm{Planck}_0 \sim 67.3\,
\hbox{km}\cdot\hbox{s}^{-1} \cdot \hbox{Mpc}^{-1}$) and the distances
derived from baryonic acoustic oscillations (BAO). In particular we
extend the exponential cutoff model used in
Refs.~\cite{PlanckCollaboration2014, Ade:2015lrj} by including a
maximum-deficit parameter, which leads us to new results as we will see
below.

In our original motivating model for a physical large-scale power deficit, the
primordial perturbations are produced in a preinflationary radiation-dominated
phase \cite{Wang:2007ws} that is in a state of ``quantum nonequilibrium''
\cite{Colin:2014pna,Colin:2015tla,Colin:2013rwa} (resulting in violations of the
usual Born rule, a possibility that is allowed in the de Broglie-Bohm pilot-wave
formulation of quantum mechanics \cite{Valentini:2008dq}). Dynamical relaxation
to quantum equilibrium (that is, to the Born rule) is found to be suppressed at
very large wavelengths, thereby naturally producing a dip in the primordial
power spectrum at large scales. If we add the simplifying assumption that the
spectrum is unchanged by the transition from preinflation to inflation, we
obtain a three-parameter modification of the CMB spectrum (noting that quantum
relaxation may be shown to not take place during inflation itself
\cite{Valentini:2008dq}). In this paper we extend the modification to four
parameters in order to be able to compare with the cases studied by the Planck
team.

We have found that Planck data (temperature only, with no polarization)
combined with the low-redshift determination of $H_0$ and the distances
derived from baryonic oscillations are able to constrain two of the new
parameters fairly well, yielding a moderate improvement at the
$2-3\sigma$ level in favor of our quantum relaxation model (in
particular for the combination of temperature data with $H_0$ alone).
However, the significance of the fits tends to decrease when
polarization data are added. We then seem driven to the conclusion that
our starting point for a modified power spectrum yields statistically
inconclusive results. Alternatively, however, peculiarities of the fit
when polarization data are added suggest that a better fit might be
obtained in a model that allows for a breaking of scalar-tensor
consistency. Our analysis suggests a motivation for considering such
models, which arise naturally in quantum relaxation scenarios. Our
results also suggest a motivation for considering models with negligible
power in the far super-Hubble limit.

Additionally, in the course of our analysis we found that the fitting
process led naturally to a preference for an extreme case of our
(scalar) deficit function. The best fit seems to be obtained with a
simple two-parameter sharp decrease in the power spectrum, with a
statistical significance ranging from substantial to strong (according
to Jeffreys' scale given in Ref.~\cite[Appendix B]{Jeffreys1998}). We
find a good account of the data with a sudden dip of about 20\% at a
characteristic scale of around $2\pi k^{-1} \approx
2200~\mathrm{Mpc}~(k^{-1} \approx 350~\mathrm{Mpc})$; a first analysis
leading to a similar effect was discussed in Ref. \cite{Hazra:2013nca}
for the Planck 2013 data. Whether or not this provides a new
physically relevant scale in other areas of cosmology is left for
future investigation.

In Sec.~\ref{sec:model} we present our cosmological model and our
parametrizations of the modified power spectrum. In Sec.~\ref{sec:meth} we
describe the methodology for our statistical data analysis. Our numerical
approach is summarized in Sec.~\ref{sec:num}. Our results are presented and
discussed in detail in Sec.~\ref{sec:results}. The significance and
properties of the sudden jump deficit function are discussed in Section
\ref{sec:jump}. A possible breaking of scalar-tensor consistency is briefly
addressed in Sec.~\ref{rTS}. The implications of our results for future work
on quantum relaxation scenarios are summarized in Sec.~\ref{sec:QR}.  Our
conclusions are drawn in Sec.~\ref{sec:conc}. These are followed by two
appendices. Appendix \ref{app:numcosmo} discusses the cosmology library used in
our numerical analysis, while Appendix \ref{App:models} provides more details of
our data analysis.

\section{Cosmological Model}
\label{sec:model}

Even though the CMB anisotropies depend strongly on the primordial power
spectrum, they also depend on other aspects of the cosmological
model which are unrelated to the origin of the primordial perturbations.
For this reason, we start by specifying the complete cosmological model
that we use to calculate the CMB anisotropies theoretically.

\subsection{Deficit functions} \label{sec:deficit}

In what follows we do not adopt a particular inflationary model (or any
reasonable alternative one might consider
\cite{Peter:2008qz,Battefeld2015, Brandenberger:2016vhg,
2018PhRvD..97h3517B}) but only a simple power-law model of the fiducial
power spectrum as best fitted by all currently available data. We also
assume, again in accordance with known data, that only the adiabatic
mode is present. We do not consider any contribution from gravitational
waves. In such a framework, all the information about the primordial
perturbations is contained in the fiducial power spectrum (with
``$_\mathrm{plaw}$'' denoting ``power-law'')
\begin{equation}
\mathcal{P}_{\mathrm{F}}(k) = \mathcal{P}_\mathrm{plaw}(k) \equiv
\mathcal{A}_\mathrm{s}\left(\frac{k}{k_\star}\right)^{n_\mathrm{s}-1},
\label{eq:PPS:fiduc}
\end{equation}
where $k_\star$ is the pivotal mode chosen (following the Planck
analysis) to be $k_\star = 0.05 \, \mathrm{Mpc}^{-1}$,
$\mathcal{A}_\mathrm{s}$ is the amplitude of the adiabatic mode measured
at $k_\star$, and $n_\mathrm{s}$ is the spectral index. The modified
power spectrum may then be described by a deficit function $\xi(k)$, or
alternatively $\chi(k) \equiv 1-\xi(k)$, defined by
\begin{equation}
\mathcal{P}(k) = \xi(k) \mathcal{P}_{\mathrm{F}}(k) = \left[
1-\chi(k)\right] \mathcal{P}_{\mathrm{F}}(k).
\label{eq:PPS}
\end{equation}
Here $\mathcal{P}(k)$ is the effective (to be estimated) power
spectrum and $\lim_{k\gg k_\mathrm{c}} \chi(k) =0$ for some physical
wave number $k_\mathrm{c}$ to be determined by the data. This
power spectrum approximates the fiducial one at small scales ($k\gg
k_\mathrm{c}$, where $\xi\to 1$) and modifies it at large scales
($k\ll k_\mathrm{c}$). Note that this approach does not model the
primordial mechanism in play \cite{Chluba:2015bqa} but merely assumes
a phenomenological form for the resulting spectrum.

As will be made explicit below, the phenomenological deficit function
depends not only on the scale $k$ but also on a set of parameters
$\theta_\xi$, so that, in principle, one should write
$\xi(k,\theta_\xi)$ instead of $\xi(k)$ in Eq.~\eqref{eq:PPS}. In order
to simplify the notation, we shall instead consider specific choices for
the deficit function, whose name then encodes the relevant set of
parameters (as defined in what follows).

References~\cite{PlanckCollaboration2014, Ade:2015lrj} considered two
phenomenological models of the CMB power deficit at low multipoles. The first is
the so-called exponential cutoff~\cite{2003JCAP...07..002C}, referred to by a
subscript $_\mathrm{expc}$ in what follows, which we modify slightly to include
the possibility of a large-scale renormalization:
\begin{equation}
\chi_\mathrm{expc}(k) = 1-\xi_\mathrm{expc}(k)  = (1-\beta)
\exp\left[-\left(\frac{k}{k_\mathrm{c}}\right)^{\lambda} \right],
\label{eq:def:expc}
\end{equation}
where $k_\mathrm{c}$ explicitly controls the cutoff wavelength that was
implicit in \eqref{eq:PPS}, $\lambda$ provides a transition rate, and $\beta$ is
introduced to mimic the large-scale behavior of our more general model (given
below) and it acts as a maximum deficit. This parametrization indeed leaves the
small scales unchanged: $\lim_{k\gg k_\mathrm{c}} \chi_\mathrm{expc}(k) =0$. For
the large-scale limit we obtain
\begin{equation*}
\xi_\mathrm{expc}(k) \underset{k \ll k_\mathrm{c}}{\approx} 
\beta + (1 -
\beta)\left(\frac{k}{k_\mathrm{c}}\right)^{\lambda},
\end{equation*}
so that for $\beta \neq 0$ the spectrum is merely rescaled by the
constant $\beta$. On the other hand, when $\beta = 0$ (as in the
original study by the Planck team) the power spectrum becomes
\begin{equation}
\mathcal{P}(k) \underset{k \ll k_\mathrm{c}}{\approx}
\left(\frac{k_\star}{k_\mathrm{c}} \right)^{\lambda} \As \left(
\frac{k}{k_\star} \right)^{\ns + \lambda - 1}.
\end{equation}
This expression adds freedom (at large scales) to the spectral index
through the parameter $\lambda$, which at the same time controls the
transition rate and the large-scale power-law behavior. In the
following we employ three different choices of parameter sets:
\textbf{expc3}, which labels the model with all the parameters
($k_\mathrm{c}$,\;$\lambda$,\;$\beta$) freely varying, \textbf{expc2}
(which coincides with the model used the Planck team) where we set
$\beta= 0$, and \textbf{expc1} where in addition to $\beta = 0$ we
impose the further constraint $\lambda = \frac12$.

The second model introduced in Ref.~\cite{Ade:2015lrj}
consists of a broken power law:
\begin{equation}
\xi_\mathrm{bpl}(k) = 
\begin{cases}
\left( \displaystyle\frac{k}{k_\mathrm{c}}\right)^{\lambda}
& \hbox{for} \ \ \ \ k \leq k_\mathrm{c}, \\ 1
& \hbox{for} \ \ \ \ k \geq k_\mathrm{c}.
\end{cases}
\label{eq:def:bpl}
\end{equation}
We shall refer to this single parametrization, with both parameters
$k_\mathrm{c}$ and $\lambda$ freely varying, as \textbf{bpl}.

\begin{figure*}
	\centering
	\includegraphics[width=0.44\textwidth,height=0.44\textwidth]{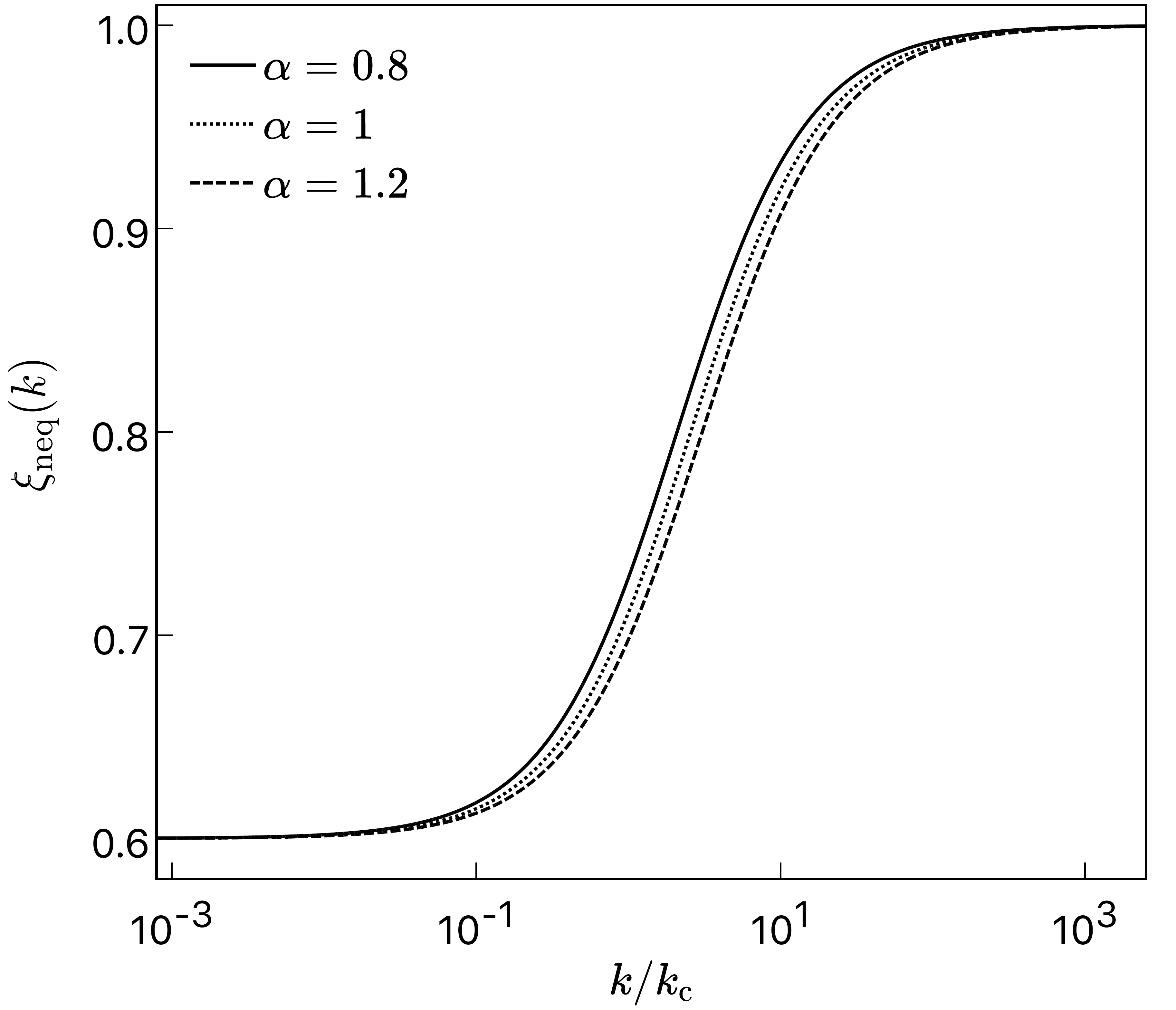}\hskip5mm
	\includegraphics[width=0.44\textwidth,height=0.44\textwidth]{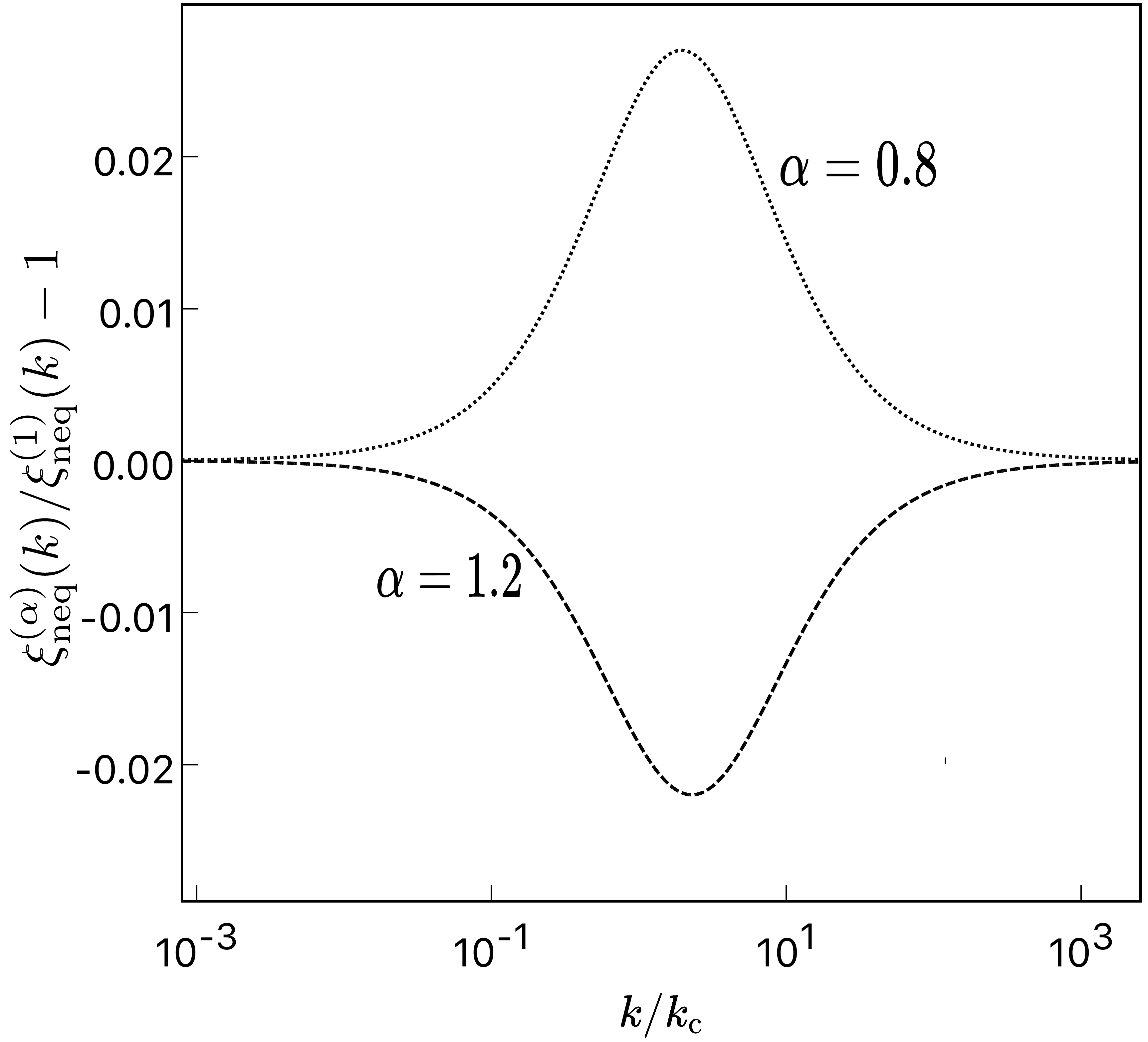}
	\caption{Effect of the parameter $\alpha$ on $\xi_\mathrm{neq}(k)$ 
	as defined through Eq.~\eqref{xineq} for $\beta=0.6$
	and $\lambda=1$. The left panel shows the function
	$\xi_\mathrm{neq}(k)$ for different values of $\alpha$ while the right
	panel exhibits the fractional difference from the fiducial with $\alpha
	=1$. For the relevant range (see the fits below), the effect is at best
	of order a few percent. \label{fig:xi_cmp}}
\end{figure*}

A more general parametrization has been obtained in the framework of
quantum nonequilibrium initial conditions \cite{Valentini:2008dq}. In
this setting one assumes that the quantum wave functional is the usual
vacuum, but the actual field variables take values whose variance is
smaller than the usual quantum variance (a feature that is possible in
the de Broglie-Bohm formulation of quantum mechanics). If some
fluctuations exit the Hubble scale while still in a nonequilibrium
state, they may be stuck with a low-variance distribution until they
become classical. To obtain a prediction, Ref.~\cite{Colin:2014pna}
considered quantum relaxation (from initial nonequilibrium) for a
spectator scalar field during a preinflationary radiation-dominated
phase and calculated the resulting power spectrum. Adding the
simplifying assumption that the spectrum is unchanged during the
transition from preinflation to inflation, a ``quantum relaxation''
deficit function $\xi_\mathrm{neq}(k)$ was found which, after
generalizing from a fixed index $\lambda = 1$ to an arbitrary index
$\lambda$, reads\footnote{To connect with the notation of
Ref.~\cite{Colin:2014pna}, the coefficient there denoted $c_{3}$ is
equal to our $1/\alpha$ while the function there denoted $\xi(k)$ is
equal to our $\xi_{\mathrm{neq}}(k)/\alpha$.}
\begin{equation}
\xi_\mathrm{neq}(k) = 1- \alpha\left\{\frac{\pi}{2} - \arctan \left[  \left(
\frac{k}{k_\mathrm{c}}\right)^\lambda  + b
\right]\right\}. \label{xineq0}
\end{equation}
The parameter $b$ is constrained by the physical requirement that
$\xi_\mathrm{neq}(k)>0$ for all $k$. In the limit $k\to 0$ this requires
$$
b > \tan\left( \frac{\pi}{2} - \frac{1}{\alpha} \right)
$$
to avoid the spectrum becoming negative. As before the parameters
$\alpha$ and $\lambda$ as well as the characteristic scale
$k_\mathrm{c}$ must be positive definite. While the latter three
constraints are mostly harmless, the condition on $b$ presents a
challenge. Dealing with such a constrained parameter space is often
impractical when performing a statistical analysis, and we therefore
avoid it by introducing a new parameter $\beta$ defined implicitly by
\begin{equation}
b\left(\alpha,\;\beta\right) \equiv \tan\left( \frac{\pi}{2} - \frac{\vert 1-\beta\vert}{\alpha} \right).
\label{bbeta}
\end{equation}
The constraint on $b$ may be recast as the simpler requirements 
$$0 < \beta < 2, \qquad 1-\alpha\pi < \beta < 1+\alpha\pi,$$
where the second one arises from the domain restriction of
$\tan$. Finally we further simplify the constraints by imposing $\alpha
> 1/\pi$, yielding the complete set
$$ 0 < \beta < 2,\quad \frac{1}{\pi}< \alpha,  \quad 0 < \lambda.$$
The quantum relaxation deficit function $\xi_\mathrm{neq}(k)$ now spans
a simpler space which does not pose any serious numerical threat. It
reads, explicitly,
\begin{align}
\chi_\mathrm{neq} & = \text{sign}(1-\beta)\alpha\left\{\frac{\pi}{2} -
\arctan \left[  \left( \frac{k}{k_\mathrm{c}}\right)^\lambda  +
b\left(\alpha,\;\beta\right) \right]\right\} \nonumber \\
&\equiv 1- \xi_\mathrm{neq}(k). \label{xineq}
\end{align}
A few examples are shown in Fig.~\ref{fig:xi_cmp}, which also emphasizes
that the parameter $\alpha$ does not play a very important role.

The deficit function \eqref{xineq} shares many properties with the
Planck-exponential and broken-power laws. To begin with, the
long-wavelength limit is
\begin{equation}
\lim_{k\to 0} \xi_\mathrm{neq}(k) = \beta,
\end{equation}
so that on large scales ($k \ll k_\mathrm{c}$) we have
\begin{equation}
\mathcal{P}(k) \underset{k \ll k_\mathrm{c}}{\approx}
\beta \As\left(\frac{k}{k_\star}\right)^{\ns -1}.
\end{equation}
Thus $\xi_\mathrm{neq}(k)$ behaves asymptotically like a step function
that modifies the spectrum by multiplying it by $\beta$ on large scales
and leaving it unchanged on small scales.

The shape of the transition itself depends on both $\alpha$ and
$\beta$, which in principle act independently. However, by plotting
$\xi_\mathrm{neq}(k)$ for different values of $\alpha$ (see
Fig.~\ref{fig:xi_cmp}) it is easy to see that varying $\alpha$ modifies
$\xi_\mathrm{neq}(k)$ only during the transition ($k \approx
k_\mathrm{c}$) and even then only slightly. Figure~\ref{fig:xi_cmp}
illustrates this for three values of $\alpha$, specifically 0.8, 1.0
and 1.2 (and with $\beta = 0.6$), leading to a difference of at most
$\approx 2\%$ in the transition region. We also investigated this
effect for other values of $\beta$. For the most extreme cases, namely
$\beta \ll 1$ and $\alpha > 1.5$ or $\alpha < 0.5$, we obtained a
maximum difference of order $50\%$ at the transition point.

The additional index $\lambda$ included in \eqref{xineq} permits two
different effects (similarly to the \textbf{expc} case). First, in the
limit $\beta\to 0$ one finds
\begin{equation}
\xi_\mathrm{neq}(k)\big|_{\beta\to 0}
\underset{k \ll k_\mathrm{c}}{\approx}
\alpha \sin^2 \left( \frac{1}{\alpha}\right)
\left(\frac{k}{k_\mathrm{c}}\right)^\lambda.
\label{xibeta0}
\end{equation}
For this choice of parameters the deficit function $\xi_\mathrm{neq}(k)$
changes the spectral index by adding $\lambda$ to the power of $k$. In
other words, for large scales $\mathcal{P}(k \to 0) \propto
k^{\ns+\lambda-1}$ and we recover the large-scale behavior of the broken
power law.

The second effect of $\lambda$ is that it controls the rate of the
transition between large and small scales, that is, how many decades it
takes for $\xi_\mathrm{neq}(k)$ to become approximately constant
(numerically saturated to machine precision) for $k > k_\mathrm{c}$ and
$k < k_\mathrm{c}$. For example, the left panel of Fig.~\ref{fig:xi_cmp}
shows that $\xi_\mathrm{neq}(k)$ is essentially constant as soon as
$k/k_c \lesssim 10^{-2}$ or $k/k_c > 10^{2}$. The greater the value of
$\lambda$, the faster the transition takes place (this transition is
discussed in detail in Appendix~\ref{app:atan}). In fact$-$and this will
play an important role in our analysis$-$in the limit $\lambda \gg 1$
and with $\beta\not= 0$ the deficit function becomes a step function and
thus induces a sharp jump in the primordial power spectrum.

In short, using the final parametrization we have the following
parameter space
\begin{equation*}
(k_\mathrm{c},\; \lambda,\; \alpha,\; \beta).
\end{equation*}
The first parameter $k_\mathrm{c}$ sets the physical scale at which the
transition occurs, $\lambda$ controls the transition rate (and if
$\beta=0$ the broken power-law behavior), $\beta$ represents the
amplitude of the drop of the power-law spectrum, and $\alpha$
parametrize the shape of the transition (although very weakly).

Given the precision with which the current data (see
Refs.\cite{PlanckCollaboration2014, Ade:2015lrj}) can
constrain the PPS, we consider three different regions of the parameter
space. First we consider the entire space, which we call {\bf atan4},
and fit all four parameters to the data (following the same nomenclature
as for the cases above). In the second parametrization {\bf atan3} we
restrict attention to the subset $\lambda = 1$, so we measure only the
position, the rescaling and the shape of the power-law modification.
Next we additionally fix $\alpha = 1$, yielding the {\bf atan2}
parametrization with the shape parameter removed. Finally, in the fourth
parametrization {\bf atan1}, we keep $\lambda=1$ and $\alpha = 1$ and
we additionally impose $\beta = 0$, thus measuring only the transition
point $k_\mathrm{c}$. This last case, as discussed above, shares many
characteristics with the broken power law.

When we consider models with some parameters fixed, this is akin to
setting a very strong prior (essentially a delta function). Strictly
speaking this should be done only in the context of a well-defined
theoretical framework. In a purely phenomenological description, it is
important to note that results for such restricted models are only
illustrative and their statistical significance should not be taken too
literally. As we shall see, some of the restricted models perform
comparatively well, but their significance might not necessarily be
physically meaningful. In general, such restricted fits merely serve to test if a given dataset is able to constrain a given parameter. From the results presented in Tables~\ref{tab:TT_onlyMP} to \ref{tab:TTTEEE_lowP_H0_BAOMP}, we see that for some data combinations all of the parameters are relevant in the sense that leaving them unconstrained improves the fit (showing that the data are sensitive to these parameters), while for other data combinations the extra parameters are irrelevant in the sense that leaving them unconstrained does not improve the fit (showing that the data are not sensitive to these parameters, or equivalently that these parameters are not constrained by the data). In an overall evaluation of the fits, we
should avoid the risk of underestimating the $p-$values by considering
$p-$values only for models with parameters that the data are actually
able to constrain (as studied in detail below).

To complete this overview of the phenomenological models
considered in the following analysis, we mention the final one appearing in
Tables~\ref{tab:TT_onlyMP} to \ref{tab:TTTEEE_lowP_H0_BAOMP}, which we have dubbed
{\bf jump}. This is a limiting case of the {\bf atan} function with the rate of transition so large ($\lambda > 40,\; \beta \neq 0$) that numerically, at double precision, the
transition appears almost discontinuous. This model was not originally included in our list of models. It was added because fitting the quantum relaxation deficit function \eqref{xineq}
with a freely-varying sharpness parameter $\lambda$ (as originally
suggested in the Planck papers \cite{PlanckCollaboration2014,Ade:2015lrj} for
the exponential cutoff and for the broken power-law) led to a value
of $\lambda$ much larger than one ($\lambda \gg 1$). Furthermore,
starting from a large value $\lambda \approx 50$ the fit was found to be stable for larger
values of $\lambda$. This led us to consider a deficit function with a sharp transition, since the function $\xi_{\mathrm{neq}}(k)$ with $\lambda \geq 50$ acts
numerically like a discontinuous jump at $k_c$. The resulting \textbf{jump} model is then
described by the following two-parameter deficit function:
\begin{equation}
\xi_\mathrm{jump}(k) = 
\begin{cases}
\beta
& \hbox{for} \ \ \ \ k \leq k_\mathrm{c}, \\ 1
& \hbox{for} \ \ \ \ k \geq k_\mathrm{c}.
\end{cases}
\label{jump}
\end{equation}
This is the deficit function employed in the tables under the label {\bf jump}.
It has some interesting properties which we discuss in Sec.~\ref{sec:jump}
below. It is worth emphasising that this new parametrization was found as an extreme
case of the \textbf{atan} and \textbf{expc} parametrizations, where the latter as originally introduced in Ref.~\cite{2003JCAP...07..002C} did not have our extra parameter
$\beta$ (where $\beta$ is necessary to obtain the \textbf{jump} model).

\subsection{$\Lambda$CDM parameters}

Apart from modifying the primordial power spectrum, our adopted cosmology is just the standard six-parameter $\Lambda$CDM model.

In the Planck analyses using C\textsc{osmo}MC~\cite{Lewis2000, Lewis2013}, the sampler
employs a parametrization with $100\theta_\mathrm{MC}$ and $\tau$ instead of
$H_0$ and $z_\mathrm{re}$, since the former are less correlated with the other
cosmological parameters. However, this is only an intermediate step. These
parameters are then converted to the ones actually used in the numerical
computation (this is explained in the C\textsc{osmo}MC
documentation\footnote{\url{https://cosmologist.info/cosmomc/readme.html}} and
can also be read directly in the code). In our analysis, both the best-fit
finder and the Markov Chain Monte Carlo (MCMC) sampler are insensitive to strong
correlations between parameters and for this reason we make direct use of the
fundamental parametrization with $H_0$ and $z_\mathrm{re}$, avoiding unnecessary
conversions between parametrizations.\footnote{Both algorithms are affine
	invariant, that is, they are invariant under linear reparametrizations. For this
	reason we were able to use the fundamental parametrization while getting a fast
	convergence of the chains. For more details on this method,
	see Refs.~\cite{Goodman2010, Foreman-Mackey2013}.} Note, however, that different
parametrizations can have a real influence on the results of an MCMC analysis if
flat priors are used. For this reason, when performing the MCMC analysis we used
two priors on the parametric space that reduce to simple flat priors for
$100\theta_\mathrm{MC}$ and $\tau$ when this parametrization is employed.

Apart from the irrelevant difference in parametrization, our cosmological model has the same ingredients as that of Ref.~\cite{Ade:2015lrj}, specifically:
\begin{itemize}
\item Hubble constant $H_0 = 100\, h\,
\hbox{km}\cdot\hbox{s}^{-1}\cdot\hbox{Mpc}^{-1}$ (thereby defining $h$).
\item Electromagnetic background radiation with a fixed temperature today
$T_{\gamma0} = 2.7255\, \hbox{K}$.
\item One massive neutrino with $m_\nu = 0.06\,\mathrm{eV}$, vanishing
chemical potential, $T_{\nu0} = 0.71611\, T_{\gamma0} $, and the
effective massless neutrino number $N_\mathrm{eff} = 2.0328$. This
configuration is such that when the massive neutrino turns
ultra-relativistic, the effective number of massless species is the
standard $3.046$. \item Cold dark matter density parametrized by
$\Omega_\mathrm{cdm}h^2$.
\item Baryon matter density parametrized by
$\Omega_\mathrm{b}h^2$.
\item Spatially flat model $\Omega_\mathcal{K} = 0$. 
\item Instantaneous reionization with
\begin{equation*}
\Delta_\mathrm{HeIII} = \Delta_\mathrm{HII} = 0.5, \quad
\lambda_\mathrm{H} = 3/2,
\end{equation*}
where $\lambda_\mathrm{H}$ is the reionization exponent and
$\Delta_\mathrm{HII}$ is the reionization width (for both HI$\to$HII and
HeI$\to$HeII) and $\Delta_\mathrm{HeIII}$ is the width for
HeII$\to$HeIII. The second reionization HeII$\to$HeIII redshift is kept
fixed at $z_\mathrm{HeIII} = 3.5$. The first reionization redshift is
employed as a free parameter $z_\mathrm{re}$. \item The fiducial PPS of
Eq.~\eqref{eq:PPS:fiduc} with the two free parameters $\As$ and $\ns$.
\item We assume negligible contributions from tensor modes. In practice
we set the tensor-to-scalar ratio $r$ to zero (on this point see,
however, Sec.~\ref{rTS}).
\end{itemize}
To summarize, our $\Lambda$CDM model depends on the free
parameters
\begin{equation}\label{eq:freeparam}
\theta_{\Lambda\mathrm{CDM}} = \{
H_0,\;\Omega_\mathrm{cdm}h^2,\;\Omega_\mathrm{b}h^2,\;z_\mathrm{re},\;
\As,\; \ns \},
\end{equation}
in addition to which one must include the PPS modification \eqref{eq:PPS} with
the choices \eqref{eq:def:expc}, \eqref{eq:def:bpl}, \eqref{xineq} or
\eqref{jump} for $\xi(k)$, thus extending the $\Lambda$CDM parameter
space $\theta_{\Lambda\mathrm{CDM}}$ by the additional $\theta_{\xi} = \{
k_\mathrm{c},\; \lambda,\; \alpha,\; \beta \}$ (depending on the case at hand).

As our last ingredient we need the Planck Foreground and Instruments
(PFI) parameters (available in Refs.~\cite{PlanckCollaboration2013b,
PlanckCollaboration2015}), yielding a final extended parameter space
\begin{equation}
\theta = \theta_{\Lambda\mathrm{CDM}}  \cup  \theta_{\xi} \cup
\theta_\mathrm{PFI}.
\label{params}
\end{equation}

\section{Methodology}\label{sec:meth}

The practical implementation of our methodology is based on specific datasets as detailed in Sec. \ref{datasets} together with a statistical analysis as detailed in Sec. \ref{statana}.

\subsection{Datasets} \label{datasets}

In our analysis we employ three different CMB datasets (with the same
nomenclature as in Ref.~\cite{PlanckCollaboration2015}). The likelihoods
are split into low-$\ell$ [for $\ell \in (2, 29)$] and high-$\ell$ (for
$\ell \geq 30$). The software adopted is the Planck likelihood code
P\textsc{lik}-2.0 (as in Ref.~\cite{PlanckCollaboration2015}), which
implements the Planck likelihood as described in
Ref.~\cite{2016A&A...594A..11P}. Our three datasets are as follows:
\begin{itemize}
	\item Planck $TT$: This refers to the low-$\ell$ and high-$\ell$
	likelihoods for CMB temperature anisotropies only (that is, for
	$C_\ell^{TT}$ only).  These two likelihoods are labeled by
	$L_\mathrm{low}$ and $L_\mathrm{high}$ respectively. The corresponding
	files for the likelihood code are:
	\begin{itemize}
		\item low-$\ell$: \textsf{commander\_rc2\_v1.1\_l2\_29\_B.clik}; 
		\item high-$\ell$: \textsf{plik\_dx11dr2\_HM\_v18\_TT.clik};
	\end{itemize}
	\item Planck $TT+$lowP: This includes the polarization data in addition
	to that of Planck $TT$ for the low-$\ell$ section, specifically
	$C_\ell^{TE}$, $C_\ell^{EE}$ and $C_\ell^{BB}$ for $\ell \in (2, 29)$.
	Note that we use the symbol $L_\mathrm{lowP}$ to refer to the
	combination of temperature and polarization for low multipoles. The
	corresponding files for the likelihood code are:
	\begin{itemize}
		\item low-$\ell$: \\
		\textsf{lowl\_SMW\_70\_dx11d\_2014\_10\_03\_v5c\_Ap.clik};
		\item high-$\ell$: \textsf{plik\_dx11dr2\_HM\_v18\_TT.clik};
	\end{itemize}
	\item Planck $TT$, $TE$, $EE$ + lowP: This includes, in addition to Planck $TT+$lowP,
	the polarization data $C_\ell^{TE}$ and $C_\ell^{EE}$ for the
	high-$\ell$ likelihood. We use the symbol $L_\mathrm{highP}$ to refer
	to this combination of temperature and polarization for high
	multipoles. The corresponding files for the likelihood code are:
	\begin{itemize}
		\item low-$\ell$: \\
		\textsf{lowl\_SMW\_70\_dx11d\_2014\_10\_03\_v5c\_Ap.clik}; 
		\item high-$\ell$: \textsf{plik\_dx11dr2\_HM\_v18\_TTTEEE.clik};
	\end{itemize}
\end{itemize}
Besides the above data likelihoods, the PFI prior is labeled by $L_\mathrm{PFI}$
(for simplicity we also use the symbol $L$ here even though this is not a
likelihood).

In addition to CMB data we also consider the 2.4\% determination of the
local value of the Hubble constant~\cite{Riess2016}. Here we use only
their best estimate $H_0 = 73.24 \pm 1.74\;
\mathrm{km\,sec^{-1}\,Mpc^{-1}}$, labeling it as H0 and its likelihood
as $L_\mathrm{H0}$. It is worth pointing out that, as discussed
in~\cite{Riess2016}, their likelihood $L(D;H_0)$ (where $D$ represent
the actual data used in~\cite{Riess2016}) is well-approximated by a
Gaussian, thus in this sense, using a Gaussian prior on $H_0$, with the
mean and variance as above, is equivalent to using their full dataset.
In our analysis we also include BAO
derived distances. Here we should stress that, differently from $H_0$, a
Gaussian likelihood on the BAO derived distances is not equivalent to
the full BAO analysis. We employ the Gaussian likelihoods on the
distances obtained from the detected BAO signals on the large-scale
correlation function as discussed in the papers below:
\begin{itemize}

\item galaxies from the 6dF Galaxy Survey (6dFGS)~\cite{Beutler2011};

\item galaxies with $z < 0.2$ from the Sloan Digital Sky Survey (SDSS)
Data Release 7 (DR7)~\cite{Ross:2014qpa};

\item galaxies from SDSS DR12 in the redshift interval $0.2 < z <
0.75$~\cite{Alam2016};

\item $147,000$ quasars from the extended Baryon Oscillation
Spectroscopic Survey (eBOSS) within $0.8 < z < 2.2$~\cite{Ata2018};

\item $137,562$ quasars with redshifts $2.1 \leq z \leq 3.5$ from the
DR11 of the BOSS/SDSS-III~\cite{Delubac2015};

\item cross-correlation of quasars with the Lyman alpha forest
absorption, using over 164,000 quasars from DR11 of the
BOSS/SDSS-III~\cite{Font-Ribera2014};

\end{itemize}
The combination of all BAO derived data is included in the likelihood
$L_\mathrm{BAO}$.\footnote{For more details on the BAO likelihoods see the data objects \texttt{NcDataBao*} at~\url{https://numcosmo.github.io/manual/ch09.html}}

\subsection{Statistical analysis} \label{statana}

In this paper we are interested in answering the following question: assuming
that the true PPS is given by $\mathcal{P}_\mathrm{F}(k)$, what is the
probability that an alternative PPS provides a better fit by pure chance? Our
strategy is to calculate this probability considering the whole fit, even though
the alternative PPS differs from the (assumed) true PPS mainly on large scales.
We use the results of the full fit to study the model quality, including small
scales and the other datasets. This avoids any kind of look-elsewhere effect and
tackles the problem in a different way. There are many works in the recent
literature modeling the large-scale behavior of the CMB anisotropies and trying
to obtain a localized signature of a physical process (see for example
Ref.~\cite{Muir:2018hjv} and references therein), whereas in our analysis we
also study the compatibility of the modifications with other datasets (as well
as their significance). 

To discriminate quickly between our models, we first address the problem
in a frequentist framework adopting the Likelihood-Ratio Test (LRT)
\cite{Craig1984, Hoel2003}. We apply this test by first identifying the
full parameter space $\theta$ (see Eq.~\eqref{params}), where each
parametrization of $\xi(k)$ introduced in Sec.~\ref{sec:model} satisfies
\begin{equation*}
\lim_{k_\mathrm{c}\to 0}\xi(k) = 1 \quad\Rightarrow\quad
\lim_{k_\mathrm{c}\to 0} \mathcal{P}(k) = \mathcal{P}_\mathrm{F}(k).
\end{equation*}
(For the \textbf{atan} and \textbf{expc} models this also occurs for
$\beta = 1$). In other words, the fiducial model is nested in the
parameter space $\theta$. In the fitting process we use the parameter
$q_\mathrm{c} = \ln(k_\mathrm{c} \times 1\,\mathrm{Mpc})$ (that is
numerically, in units of inverse Mpc) instead of $k_\mathrm{c}$. This
speeds up the numerical fitting process since the parameter
$k_\mathrm{c}$ can vary by orders of magnitude in a fit. Furthermore,
for any value\footnote{Here the speed of light $c$ enters the numerical
analysis (given the units used).} $q_\mathrm{c} \ll \ln (H_0 / c \times
1\,\mathrm{Mpc})$ the deficit function $\xi(k)$ is close to one in the
whole physical range of $k$ that influences the CMB anisotropies and is
therefore numerically indistinguishable from being taken as exactly one.
Thus the fiducial model corresponds to any value of $q_\mathrm{c} \ll
\ln (H_0 / c \times 1\,\mathrm{Mpc})$. The maximum likelihood estimator
(MLE) for $\theta$ is given by
\begin{equation*}\label{MLEtheta}
\hat{\theta} = \min_{\theta}\left\{-2\ln\left[L(D|\theta)\right]\right\}, 
\end{equation*}
where $D$ represents the dataset to be used (in our case Planck $TT$, Planck
$TT+$lowP or Planck $TT$, $TE$, $EE$+lowP and their combinations with H0, BAO
and H0+BAO), so that $L(D|\theta)$ is given by the appropriate product of
\begin{equation*}
 L_\mathrm{low},\;L_\mathrm{high},\;
 L_\mathrm{lowP},\;L_\mathrm{highP},\;L_\mathrm{PFI},\;
 L_\mathrm{H0},\;L_\mathrm{BAO}.
\end{equation*}

On the other hand the MLE for the fiducial subspace is given by
\begin{equation}\label{MLEthetaF}
\hat{\theta}_\mathrm{F} =
\min_{\theta_\mathrm{F}}\left\{-2\ln\left[L(D|\theta_\mathrm{F})\right]
\right\}, \quad \theta_\mathrm{F} \equiv
\theta_{\Lambda\mathrm{CDM}} \cup \theta_\mathrm{PFI}.
\end{equation} 
We then introduce the LRT statistic
\begin{equation}\label{LRTS}
\Gamma \equiv
-2\ln\left[\frac{L(D|\hat\theta_\mathrm{F})}{L(D|\hat\theta)}\right].
\end{equation}
It is easy to convince oneself that $\Gamma \geq 0$ since
$L(D|\hat\theta) \geq L(D|\hat\theta_\mathrm{F})$. To better understand
the effects of $\xi$ in each likelihood we also define the individual
ratios
\begin{equation}
\Gamma_i \equiv
-2\ln\left[\frac{L_i(\hat\theta_\mathrm{F})}{L_i(\hat\theta)}\right],
\end{equation}
where $i$ denotes any of (low, lowP, high, highP, BAO, H0, PFI).

In principle if we know $P(\Gamma)$ -- the probability distribution of
$\Gamma$ -- then all we need is to find $\hat{\theta}_\mathrm{F}$ and
$\hat{\theta}$ in order to compute the probability of obtaining a better
fit of $D$ in $\theta$ by chance. This probability is simply given by
\begin{equation}\label{gamma}
\gamma = \int_\Gamma^{\infty}\mathrm{d}\Gamma^\prime
P(\Gamma^\prime).
\end{equation}
Note that we choose the right-hand tail since this corresponds to a
$\hat\theta$ where the data is more probable than for
$\hat{\theta}_\mathrm{F}$. In practice, unfortunately, $P(\Gamma)$ is
not known and is hardly calculable as it would be impractical to obtain
it from first principles given the complexity of the data likelihood.
For this reason we must rely on Wilks' theorem, which asserts that in
the large-sample limit $\Gamma$ asymptotically follows a $\chi^2_r$
distribution (for a proof see Ref.~\cite{Vaart2000}), where $r$ is the
difference in dimensionality between $\theta$ and $\theta_\mathrm{F}$.
Wilks' theorem requires the fiducial model to be contained within the
parameter space. This is satisfied by the parameter $q_\mathrm{c}$
discussed above since the fiducial model requires only that
$q_\mathrm{c} \ll \ln (H_0 / c \times 1\,\mathrm{Mpc})$ and/or $\beta =
1$. In our parametrizations, the number of free parameters of  $\xi(k)$
is 4. We list the critical value of $\Gamma$ corresponding to a
$2\sigma$ probability (that is, $95.45\%$) for the relevant cases:
\begin{equation}
\chi^2_1 = 4, \quad \chi^2_2 = 6.18, \quad \chi^2_3 = 8.02, \quad
\chi^2_4 = 9.72.
\label{chi2}
\end{equation}
In other words, with one extra parameter ($r=1$) a fit giving $\Gamma >4$ means
that the fiducial model could only have generated a dataset giving this value
of $\Gamma$ (or worse) with a probability below $4.55\%$, and so on for more
parameters.

It is worth noting that the LRT has some specific features which are of
interest in our case. First, it does not depend on the choice of
variables to describe the parameter space $\theta$. Second, it naturally
takes into account the difference in the number of parameters when
comparing two nested models (see~\cite{Vaart2000}). An important caveat
when using the $\chi^2_r$ distribution is that it relies on the
asymptotic properties of the LRT.

Another aspect of the LRT worth mentioning is that it controls the
type-I error, that is, cases where the fiducial model is true but is
found to be false. For our choice of $2\sigma$ we would make a type-I
error $4.55\%$ of the time. However there is also the type-II error,
that is, cases where the alternative model is true but is found to be
false. Unlike for the type-I error, the LRT does not provide a simple
way to calculate the probability of a type-II error. We could derive
this probability analytically if the likelihood was sufficiently simple.
But in practice the likelihood is too complicated and one must resort to
simulations. A set of simulations of the alternative model must then be
produced and, for each simulation, a value of $\{\Gamma_n\}$ must be
calculated. Using this empirical distribution of $\Gamma$, one can then
calculate the probability of a type-II error. In this work we do not
address this point. But it is important to bear in mind the possibility
that, if a very small critical region is required, one may be
significantly increasing the type-II error. This would be the case, for
example, if one were requiring $5\sigma$ instead of $2\sigma$.

\begin{figure*}[ht!]
	\centering
	\includegraphics[scale=1.0]{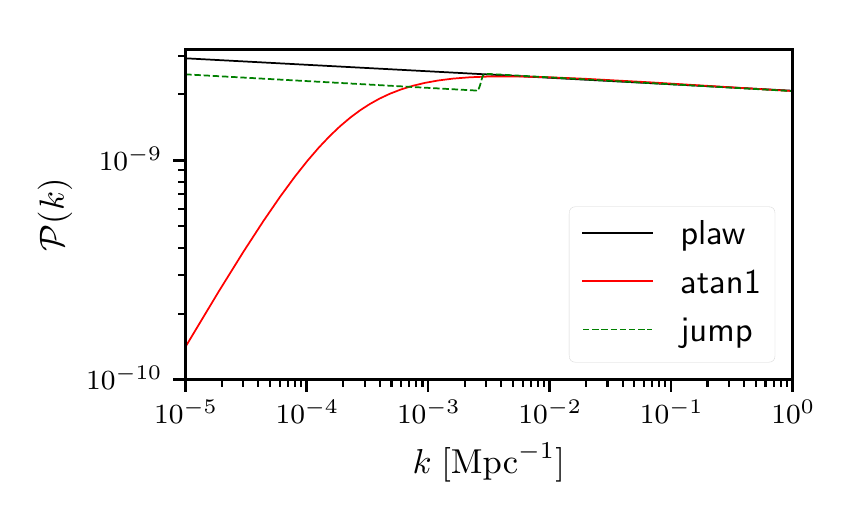}
	\includegraphics[scale=1.0]{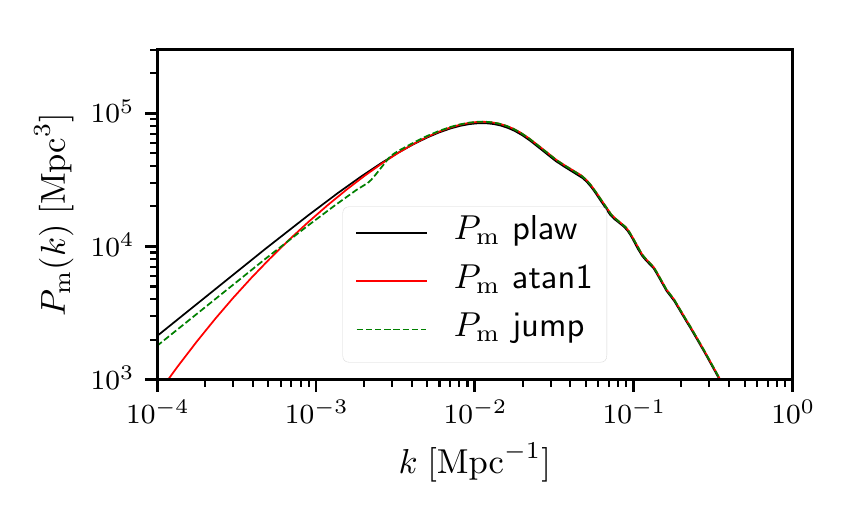}
	\caption{Upper panel: best-fit power spectra with the most successful
		deficit functions (see text). It is clear that for a smooth transition
		the relevant scale at which the deficit becomes significant is of order
		the Hubble radius, whereas in the almost discontinuous case (with a
		large value of $\lambda$) the transition scale is an order of magnitude
		smaller. Lower panel: the corresponding matter power spectra. Only the
		case of a sharp transition remains close to the case of a fiducial
		spectrum. It is an open question to understand if and how such a slight
		difference could be observable in future data. These fits are for the
		full Planck dataset. \label{fig:Pk}}
\end{figure*}

As is clear from the above discussion, the difference in the number of
parameters between the fiducial and alternative models is crucial in
determining the significance of the result [see Eq.~\eqref{chi2}].
Moreover, the use of a $\chi^2_r$ distribution is based on the
large-sample asymptotic limit. For this reason we note that, if a
parameter controls a region of the model where there are almost no data,
then we do not expect the asymptotic regime to be attained. For example,
consider the parameter $\alpha$ discussed in Sec.~\ref{sec:model}. It
modifies the PPS only in a rather narrow band of $k$ around
$k_\mathrm{c}$, and it also modifies the PPS only slightly at this
point. Consequently, we expect this parameter to be very degenerate and
not to contribute much to the fit. In an extreme case where the
alternative model has a parameter that does not modify the PPS fit at
all, it is reasonable to assume that the current data are not able to
shed any light on that aspect of the model. In such a case we also
perform the statistical tests with the degenerate parameter removed from
the analysis (keeping it fixed at some fiducial value), and we do not
take that parameter into account when comparing the alternative and
fiducial models. This ambiguity in the number of parameters is a natural
feature of our phenomenological approach. Thus the two relevant
questions are: what kind of modification of the fiducial model is the
data able to fit, and what is the significance of this fit?


\begin{table*}[ht]
  \begin{ruledtabular}
  \begin{tabular}{ c | c | c | c | c | c | c | c | c | c | c | c | r }
  $\xi(k)$ & $k_\mathrm{c}^{-1} [\mathrm{Gpc}]$ & $\lambda$ & $\beta$ & $\alpha$ & $h$ & $\ns$ & $\ln(10^{10}\As)$ & $\tau$ & $\Gamma$ & $\Gamma_\mathrm{low}$ & $\Gamma_\mathrm{high}$ & $\gamma$ \\
\hline \textbf{ plaw} &    --    &    --    &    --    &    --    & $ 0.69 $ & $ 0.98 $ & $ 3.2 $ & $ 0.15 $ & $ 0 $ & $ 0 $ & $0 $ &        --      \\
\hline \textbf{  bpl} & $ 2.9 $ & $ 7.9 $ &    --    &    --    & $ 0.69 $ & $ 0.98 $ & $ 3.2 $ & $ 0.14 $ & $ 2.7 $ & $ 2.9 $ & $-0.04 $ & $ 26 $\% ($ 1.1 \sigma$) $[2]$\\
\hline \textbf{atan1} & $ 3.6 $ & $ 1 $ & $ 0 $ & $ 1 $ & $ 0.7 $ & $ 0.97 $ & $ 3.3 $ & $ 0.18 $ & $ 5.5 $ & $ 4.4 $ & $1 $ & $  1.9 $\% ($ 2.4 \sigma$) $[1]$\\
\hline \textbf{atan2} & $ 1.4 $ & $ 1 $ & $ 0.49 $ & $ 1 $ & $ 0.7 $ & $ 0.97 $ & $ 3.4 $ & $ 0.22 $ & $ 5.8 $ & $ 5.2 $ & $0.44 $ & $  5.5 $\% ($ 1.9 \sigma$) $[2]$\\
\hline \textbf{atan3} & $ 1.7 $ & $ 1 $ & $ 0.47 $ & $ 0.94 $ & $ 0.7 $ & $ 0.97 $ & $ 3.3 $ & $ 0.2 $ & $ 5.9 $ & $ 5 $ & $0.7 $ & $ 12 $\% ($ 1.6 \sigma$) $[3]$\\
\hline \textbf{atan4} & $ 0.36 $ & $ 41 $ & $ 0.79 $ & $ 0.58 $ & $ 0.69 $ & $ 0.98 $ & $ 3.3 $ & $ 0.16 $ & $ 6.9 $ & $ 6.3 $ & $0.57 $ & $ 14 $\% ($ 1.5 \sigma$) $[4]$\\
\hline \textbf{expc1} & $ 2.1 $ & $ 0.5 $ & $ 0 $ &    --    & $ 0.69 $ & $ 0.98 $ & $ 3.3 $ & $ 0.18 $ & $ 5.6 $ & $ 4.7 $ & $0.8 $ & $  1.8 $\% ($ 2.4 \sigma$) $[1]$\\
\hline \textbf{expc2} & $ 2.7 $ & $ 0.44 $ & $ 0 $ &    --    & $ 0.69 $ & $ 0.98 $ & $ 3.3 $ & $ 0.17 $ & $ 5.8 $ & $ 4.9 $ & $0.83 $ & $  5.4 $\% ($ 1.9 \sigma$) $[2]$\\
\hline \textbf{expc3} & $ 0.33 $ & $ 14 $ & $ 0.78 $ &    --    & $ 0.69 $ & $ 0.98 $ & $ 3.3 $ & $ 0.17 $ & $ 6.7 $ & $ 6.2 $ & $0.43 $ & $  8.1 $\% ($ 1.7 \sigma$) $[3]$\\
\hline \textbf{ jump} & $ 0.36 $ &    --    & $ 0.78 $ &    --    & $ 0.69 $ & $ 0.98 $ & $ 3.3 $ & $ 0.17 $ & $ 6.8 $ & $ 6.3 $ & $0.34 $ & $  3.4 $\% ($ 2.1 \sigma$) $[2]$\\
\end{tabular}
\end{ruledtabular}
   \caption{\label{tab:TT_onlyMP} Best-fits and their significance obtained
         with the Planck $TT$ dataset. For each model we present the best-fit values of
         the cutoff scale $k_\mathrm{c}^{-1}$ in Gpc and the values of the
         dimensionless parameters $\lambda$, $\beta$ and $\alpha$. We also include the
         best-fit values of some of the $\Lambda$CDM parameters, namely, the
         dimensionless rescaled Hubble scale $h$, the spectral index $n_\mathrm{s}$ and
         amplitude $A_\mathrm{s}$ of the fiducial PPS \eqref{eq:PPS:fiduc}, and finally
         the reionization optical depth $\tau$ derived from the best-fit. We present
         the significance in two ways: the LRT statistics $\Gamma$ defined by
         \eqref{LRTS} (including the individual $\Gamma_i$ defined for each likelihood
         included in the final one) and (in the last column) the $p-$value $\gamma$
         defined by \eqref{gamma} (where we also include the probability $\gamma$
         translated into the corresponding number of one-dimensional Gaussian standard
         deviations). } 
\end{table*}


\begin{table*}[ht]
  \begin{ruledtabular}
  \begin{tabular}{ c | c | c | c | c | c | c | c | c | c | c | c | c | r }
  $\xi(k)$ & $k_\mathrm{c}^{-1} [\mathrm{Gpc}]$ & $\lambda$ & $\beta$ & $\alpha$ & $h$ & $\ns$ & $\ln(10^{10}\As)$ & $\tau$ & $\Gamma$ & $\Gamma_\mathrm{low}$ & $\Gamma_\mathrm{high}$ & $\Gamma_{H0}$ & $\gamma$ \\
\hline \textbf{ plaw} &    --    &    --    &    --    &    --    & $ 0.7 $ & $ 0.98 $ & $ 3.2 $ & $ 0.16 $ & $ 0 $ & $ 0 $ & $0 $ & $0 $ &        --      \\
\hline \textbf{  bpl} & $ 2.8 $ & $ 7.6 $ &    --    &    --    & $ 0.7 $ & $ 0.99 $ & $ 3.2 $ & $ 0.16 $ & $ 2.6 $ & $ 2.3 $ & $0.03 $ & $0.3 $ & $ 27 $\% ($ 1.1 \sigma$) $[2]$\\
\hline \textbf{atan1} & $ 3 $ & $ 1 $ & $ 0 $ & $ 1 $ & $ 0.71 $ & $ 0.98 $ & $ 3.4 $ & $ 0.21 $ & $ 7.1 $ & $ 3.8 $ & $1.8 $ & $1.4 $ & $  0.79 $\% ($ 2.7 \sigma$) $[1]$\\
\hline \textbf{atan2} & $ 1.3 $ & $ 1 $ & $ 0.5 $ & $ 1 $ & $ 0.71 $ & $ 0.97 $ & $ 3.4 $ & $ 0.23 $ & $ 8.3 $ & $ 4.8 $ & $1.6 $ & $1.8 $ & $  1.6 $\% ($ 2.4 \sigma$) $[2]$\\
\hline \textbf{atan3} & $ 1.3 $ & $ 1 $ & $ 0.51 $ & $ 1 $ & $ 0.71 $ & $ 0.97 $ & $ 3.4 $ & $ 0.23 $ & $ 8.2 $ & $ 4.8 $ & $1.6 $ & $1.8 $ & $  4.1 $\% ($ 2 \sigma$) $[3]$\\
\hline \textbf{atan4} & $ 1.3 $ & $ 1 $ & $ 0.51 $ & $ 0.99 $ & $ 0.71 $ & $ 0.97 $ & $ 3.4 $ & $ 0.23 $ & $ 8.3 $ & $ 4.8 $ & $1.6 $ & $1.8 $ & $  8.1 $\% ($ 1.8 \sigma$) $[4]$\\
\hline \textbf{expc1} & $ 1.5 $ & $ 0.5 $ & $ 0 $ &    --    & $ 0.71 $ & $ 0.99 $ & $ 3.4 $ & $ 0.22 $ & $ 6.9 $ & $ 4.3 $ & $1 $ & $1.5 $ & $  0.85 $\% ($ 2.6 \sigma$) $[1]$\\
\hline \textbf{expc2} & $ 1.6 $ & $ 0.36 $ & $ 0 $ &    --    & $ 0.72 $ & $ 0.98 $ & $ 3.4 $ & $ 0.24 $ & $ 8.6 $ & $ 4.8 $ & $1.4 $ & $2.2 $ & $  1.4 $\% ($ 2.5 \sigma$) $[2]$\\
\hline \textbf{expc3} & $ 1.6 $ & $ 0.36 $ & $ 0.007 $ &    --    & $ 0.72 $ & $ 0.98 $ & $ 3.4 $ & $ 0.24 $ & $ 8.6 $ & $ 4.8 $ & $1.4 $ & $2.2 $ & $  3.6 $\% ($ 2.1 \sigma$) $[3]$\\
\hline \textbf{ jump} & $ 0.35 $ &    --    & $ 0.76  $ &    --    & $ 0.7 $ & $ 0.99 $ & $ 3.3 $ & $ 0.2 $ & $ 7.2 $ & $ 5.9 $ & $0.76$ & $0.57 $ & $  2.8 $\% ($ 2.2 \sigma$) $[2]$\\
\end{tabular}
\end{ruledtabular}
  \caption{\label{tab:TT_only_H0MP} The same as Table~\ref{tab:TT_onlyMP} but with Planck $TT$ + H0.}
\end{table*}


\begin{table*}[ht]
  \begin{ruledtabular}
  \begin{tabular}{ c | c | c | c | c | c | c | c | c | c | c | c | c | r }
  $\xi(k)$ & $k_\mathrm{c}^{-1} [\mathrm{Gpc}]$ & $\lambda$ & $\beta$ & $\alpha$ & $h$ & $\ns$ & $\ln(10^{10}\As)$ & $\tau$ & $\Gamma$ & $\Gamma_\mathrm{low}$ & $\Gamma_\mathrm{high}$ & $\Gamma_\mathrm{BAO}$ & $\gamma$ \\
\hline \textbf{ plaw} &    --    &    --    &    --    &    --    & $ 0.68 $ & $ 0.97 $ & $ 3.2 $ & $ 0.14 $ & $ 0 $ & $ 0 $ & $0 $ & $ 0 $ &        --      \\
\hline \textbf{  bpl} & $ 2.8 $ & $ 8 $ &    --    &    --    & $ 0.68 $ & $ 0.97 $ & $ 3.2 $ & $ 0.13  $ & $ 2.6 $ & $ 2.8 $ & $-0.13$ & $-0.01 $ & $ 27 $\% ($ 1.1 \sigma$) $[2]$\\
\hline \textbf{atan1} & $ 3.5 $ & $ 1 $ & $ 0 $ & $ 1 $ & $ 0.69 $ & $ 0.97 $ & $ 3.3 $ & $ 0.18  $ & $ 4.6 $ & $ 4.1 $ & $0.41 $ & $-0.18$ & $  3.2 $\% ($ 2.1 \sigma$) $[1]$\\
\hline \textbf{atan2} & $ 2 $ & $ 1 $ & $ 0.48 $ & $ 1 $ & $ 0.69 $ & $ 0.96 $ & $ 3.3 $ & $ 0.18  $ & $ 5 $ & $ 5 $ & $0.09 $ & $-0.18$ & $  8.1 $\% ($ 1.7 \sigma$) $[2]$\\
\hline \textbf{atan3} & $ 2 $ & $ 1 $ & $ 0.48 $ & $ 1 $ & $ 0.69 $ & $ 0.96 $ & $ 3.3 $ & $ 0.18  $ & $ 5 $ & $ 5 $ & $0.09 $ & $-0.18$ & $ 17 $\% ($ 1.4 \sigma$) $[3]$\\
\hline \textbf{atan4} & $ 0.35 $ & $ 48 $ & $ 0.77 $ & $ 0.51 $ & $ 0.69 $ & $ 0.98 $ & $ 3.3 $ & $ 0.18 $ & $ 5.5 $ & $ 6.5 $ & $-0.16 $ & $-1.1 $ & $ 24 $\% ($ 1.2 \sigma$) $[4]$\\
\hline \textbf{expc1} & $ 2.2 $ & $ 0.5 $ & $ 0 $ &    --    & $ 0.69 $ & $ 0.97 $ & $ 3.3 $ & $ 0.16 $ & $ 5.3 $ & $ 4.7 $ & $0.52 $ & $-0.02 $ & $  2.1 $\% ($ 2.3 \sigma$) $[1]$\\
\hline \textbf{expc2} & $ 2.5 $ & $ 0.45 $ & $ 0 $ &    --    & $ 0.69 $ & $ 0.97 $ & $ 3.3 $ & $ 0.17 $ & $ 5.5 $ & $ 4.9 $ & $0.46 $ & $ 0.02 $ & $  6.4 $\% ($ 1.9 \sigma$) $[2]$\\
\hline \textbf{expc3} & $ 0.33 $ & $ 20 $ & $ 0.79 $ &    --    & $ 0.69 $ & $ 0.97 $ & $ 3.3 $ & $ 0.16 $ & $ 6.7 $ & $ 6.4 $ & $0.14 $ & $-0.02 $ & $  8.3 $\% ($ 1.7 \sigma$) $[3]$\\
\hline \textbf{ jump} & $ 0.36 $ &    --    & $ 0.79 $ &    --    & $ 0.68 $ & $ 0.97 $ & $ 3.3 $ & $ 0.16 $ & $ 6.8 $ & $ 6.4 $ & $0.22 $ & $ 0.03 $ & $  3.4 $\% ($ 2.1 \sigma$) $[2]$\\
\end{tabular}
\end{ruledtabular}
  \caption{\label{tab:TT_only_BAOMP} The same as Table~\ref{tab:TT_onlyMP} but with Planck $TT$ + BAO.}
\end{table*}


\begin{table*}[ht]
  \begin{ruledtabular}
  \begin{tabular}{ c | c | c | c | c | c | c | c | c | c | c | c | c | c | r }
  $\xi(k)$ & $k_\mathrm{c}^{-1} [\mathrm{Gpc}]$ & $\lambda$ & $\beta$ & $\alpha$ & $h$ & $\ns$ & $\ln(10^{10}\As)$ & $\tau$ & $\Gamma$ & $\Gamma_\mathrm{low}$ & $\Gamma_\mathrm{high}$ & $\Gamma_{H0}$ & $\Gamma_\mathrm{BAO}$ & $\gamma$ \\
\hline \textbf{ plaw} &    --    &    --    &    --    &    --    & $ 0.69 $ & $ 0.98 $ & $ 3.2 $ & $ 0.13 $ & $ 0 $ & $ 0 $ & $0 $ & $0 $ & $ 0 $ &        --      \\
\hline \textbf{  bpl} & $ 3 $ & $ 7.9 $ &    --    &    --    & $ 0.69 $ & $ 0.98 $ & $ 3.2 $ & $ 0.14 $ & $ 2.6 $ & $ 2.2 $ & $0.34 $ & $0.08 $ & $-0.05 $ & $ 27 $\% ($ 1.1 \sigma$) $[2]$\\
\hline \textbf{atan1} & $ 3.9 $ & $ 1 $ & $ 0 $ & $ 1 $ & $ 0.69 $ & $ 0.97 $ & $ 3.3 $ & $ 0.17 $ & $ 5.6 $ & $ 3.5 $ & $1.6 $ & $0.8 $ & $-0.53 $ & $  1.8 $\% ($ 2.4 \sigma$) $[1]$\\
\hline \textbf{atan2} & $ 2.6 $ & $ 1 $ & $ 0.49 $ & $ 1 $ & $ 0.69 $ & $ 0.97 $ & $ 3.3 $ & $ 0.17 $ & $ 5.8 $ & $ 4 $ & $1.3 $ & $0.94 $ & $-0.62 $ & $  5.6 $\% ($ 1.9 \sigma$) $[2]$\\
\hline \textbf{atan3} & $ 1.4 $ & $ 1 $ & $ 0.5 $ & $ 0.54 $ & $ 0.69 $ & $ 0.97 $ & $ 3.3 $ & $ 0.17 $ & $ 5.8 $ & $ 4 $ & $1.4 $ & $0.79 $ & $-0.5 $ & $ 12 $\% ($ 1.5 \sigma$) $[3]$\\
\hline \textbf{atan4} & $ 0.35 $ & $ 46 $ & $ 0.77 $ & $ 0.52 $ & $ 0.69 $ & $ 0.98 $ & $ 3.3 $ & $ 0.18 $ & $ 6.5 $ & $ 5.6 $ & $0.58 $ & $0.63 $ & $-0.46 $ & $ 16 $\% ($ 1.4 \sigma$) $[4]$\\
\hline \textbf{expc1} & $ 2.1 $ & $ 0.5 $ & $ 0 $ &    --    & $ 0.69 $ & $ 0.98 $ & $ 3.3 $ & $ 0.18 $ & $ 5.8 $ & $ 3.8 $ & $1.4 $ & $0.63 $ & $-0.37 $ & $  1.6 $\% ($ 2.4 \sigma$) $[1]$\\
\hline \textbf{expc2} & $ 2.5 $ & $ 0.44 $ & $ 0 $ &    --    & $ 0.69 $ & $ 0.98 $ & $ 3.3 $ & $ 0.18 $ & $ 6 $ & $ 4.1 $ & $1.4 $ & $0.48 $ & $-0.2 $ & $  4.9 $\% ($ 2 \sigma$) $[2]$\\
\hline \textbf{expc3} & $ 0.33 $ & $ 17 $ & $ 0.77 $ &    --    & $ 0.69 $ & $ 0.98 $ & $ 3.3 $ & $ 0.18 $ & $ 6.5 $ & $ 5.5 $ & $0.7 $ & $0.55 $ & $-0.38 $ & $  8.8 $\% ($ 1.7 \sigma$) $[3]$\\
\hline \textbf{ jump} & $ 0.35 $ &    --    & $ 0.77 $ &    --    & $ 0.69 $ & $ 0.98 $ & $ 3.3 $ & $ 0.18 $ & $ 6.7 $ & $ 5.6 $ & $0.76 $ & $0.31 $ & $-0.14 $ & $  3.5 $\% ($ 2.1 \sigma$) $[2]$\\
\end{tabular}
\end{ruledtabular}
  \caption{\label{tab:TT_only_H0_BAOMP} The same as Table~\ref{tab:TT_onlyMP} but with Planck $TT$ + H0 + BAO.}
\end{table*}
 

\begin{table*}
  \begin{ruledtabular}
  \begin{tabular}{ c | c | c | c | c | c | c | c | c | c | c | c | r }
  $\xi(k)$ & $k_\mathrm{c}^{-1} [\mathrm{Gpc}]$ & $\lambda$ & $\beta$ & $\alpha$ & $h$ & $\ns$ & $\ln(10^{10}\As)$ & $\tau$ & $\Gamma$ & $\Gamma_\mathrm{low}$ & $\Gamma_\mathrm{high}$ & $\gamma$ \\
\hline \textbf{ plaw} &    --    &    --    &    --    &    --    & $ 0.68 $ & $ 0.97 $ & $ 3.1 $ & $ 0.085 $ & $ 0 $ & $ 0 $ & $0 $ &        --      \\
\hline \textbf{  bpl} & $ 1.2 $ & $ 0.68 $ &    --    &    --    & $ 0.68 $ & $ 0.97 $ & $ 3.1 $ & $ 0.1 $ & $ 2.1 $ & $ 0.9 $ & $1.1 $ & $ 34 $\% ($ 0.95 \sigma$) $[2]$\\
\hline \textbf{atan1} & $ 5.3 $ & $ 1 $ & $ 0 $ & $ 1 $ & $ 0.68 $ & $ 0.96 $ & $ 3.2 $ & $ 0.11 $ & $ 3.5 $ & $ 2.3 $ & $1.1 $ & $  6.2 $\% ($ 1.9 \sigma$) $[1]$\\
\hline \textbf{atan2} & $ 5.2 $ & $ 1 $ & $ 0.01 $ & $ 1 $ & $ 0.68 $ & $ 0.96 $ & $ 3.1 $ & $ 0.1 $ & $ 3.5 $ & $ 2.4 $ & $1.1 $ & $ 17 $\% ($ 1.4 \sigma$) $[2]$\\
\hline \textbf{atan3} & $ 3.7 $ & $ 1 $ & $ 0.009 $ & $ 0.52 $ & $ 0.68 $ & $ 0.96 $ & $ 3.1 $ & $ 0.1 $ & $ 3.6 $ & $ 2.1 $ & $1.4 $ & $ 30 $\% ($ 1 \sigma$) $[3]$\\
\hline \textbf{atan4} & $ 0.38 $ & $ 37 $ & $ 0.81 $ & $ 0.51 $ & $ 0.68 $ & $ 0.97 $ & $ 3.1 $ & $ 0.11 $ & $ 5 $ & $ 3 $ & $1.9 $ & $ 29 $\% ($ 1.1 \sigma$) $[4]$\\
\hline \textbf{expc1} & $ 2.9 $ & $ 0.5 $ & $ 0 $ &    --    & $ 0.68 $ & $ 0.97 $ & $ 3.2 $ & $ 0.11 $ & $ 3.9 $ & $ 1.8 $ & $1.9 $ & $  5 $\% ($ 2 \sigma$) $[1]$\\
\hline \textbf{expc2} & $ 2.7 $ & $ 0.52 $ & $ 0 $ &    --    & $ 0.68 $ & $ 0.97 $ & $ 3.2 $ & $ 0.11 $ & $ 3.9 $ & $ 1.8 $ & $1.8 $ & $ 14 $\% ($ 1.5 \sigma$) $[2]$\\
\hline \textbf{expc3} & $ 0.35 $ & $ 14 $ & $ 0.8 $ &    --    & $ 0.68 $ & $ 0.97 $ & $ 3.2 $ & $ 0.11 $ & $ 4.7 $ & $ 2.5 $ & $1.9 $ & $ 19 $\% ($ 1.3 \sigma$) $[3]$\\
\hline \textbf{ jump} & $ 0.37 $ &    --    & $ 0.81 $ &    --    & $ 0.68 $ & $ 0.97 $ & $ 3.1 $ & $ 0.1 $ & $ 5.1 $ & $ 3.1 $ & $1.7 $ & $  8 $\% ($ 1.8 \sigma$) $[2]$\\
\end{tabular}
\end{ruledtabular}
  \caption{\label{tab:TT_lowPMP} The same as Table~\ref{tab:TT_onlyMP} but with Planck $TT$ + lowP.}
\end{table*}


\begin{table*}
  \begin{ruledtabular}
  \begin{tabular}{ c | c | c | c | c | c | c | c | c | c | c | c | c | r }
  $\xi(k)$ & $k_\mathrm{c}^{-1} [\mathrm{Gpc}]$ & $\lambda$ & $\beta$ & $\alpha$ & $h$ & $\ns$ & $\ln(10^{10}\As)$ & $\tau$ & $\Gamma$ & $\Gamma_\mathrm{lowP}$ & $\Gamma_\mathrm{high}$ & $\Gamma_{H0}$ & $\gamma$ \\
\hline \textbf{ plaw} &    --    &    --    &    --    &    --    & $ 0.69 $ & $ 0.97 $ & $ 3.1 $ & $ 0.093 $ & $ 0 $ & $ 0 $ & $0 $ & $0 $ &        --      \\
\hline \textbf{  bpl} & $ 1.2 $ & $ 0.71 $ &    --    &    --    & $ 0.69 $ & $ 0.98 $ & $ 3.1 $ & $ 0.11 $ & $ 2.9 $ & $ 1 $ & $0.85 $ & $0.89 $ & $ 23 $\% ($ 1.2 \sigma$) $[2]$\\
\hline \textbf{atan1} & $ 5.6 $ & $ 1 $ & $ 0 $ & $ 1 $ & $ 0.69 $ & $ 0.97 $ & $ 3.2 $ & $ 0.11 $ & $ 3.8 $ & $ 1.6 $ & $1.6 $ & $0.49 $ & $  5.1 $\% ($ 1.9 \sigma$) $[1]$\\
\hline \textbf{atan2} & $ 5.4 $ & $ 1 $ & $ 0.007 $ & $ 1 $ & $ 0.69 $ & $ 0.97 $ & $ 3.2 $ & $ 0.11 $ & $ 3.9 $ & $ 1.6 $ & $1.8 $ & $0.38 $ & $ 15 $\% ($ 1.5 \sigma$) $[2]$\\
\hline \textbf{atan3} & $ 3.6 $ & $ 1 $ & $ 0.003 $ & $ 0.5 $ & $ 0.69 $ & $ 0.97 $ & $ 3.2 $ & $ 0.12 $ & $ 4.1 $ & $ 1 $ & $2.4 $ & $0.5 $ & $ 25 $\% ($ 1.2 \sigma$) $[3]$\\
\hline \textbf{atan4} & $ 0.38 $ & $ 41 $ & $ 0.81 $ & $ 0.56 $ & $ 0.69 $ & $ 0.97 $ & $ 3.2 $ & $ 0.11 $ & $ 4.7 $ & $ 2 $ & $3.2 $ & $-0.68 $ & $ 32 $\% ($ 0.99 \sigma$) $[4]$\\
\hline \textbf{expc1} & $ 3.2 $ & $ 0.5 $ & $ 0 $ &    --    & $ 0.69 $ & $ 0.97 $ & $ 3.2 $ & $ 0.12 $ & $ 4 $ & $ 1.1 $ & $2.3 $ & $0.3 $ & $  4.6 $\% ($ 2 \sigma$) $[1]$\\
\hline \textbf{expc2} & $ 2.7 $ & $ 0.57 $ & $ 0 $ &    --    & $ 0.69 $ & $ 0.97 $ & $ 3.2 $ & $ 0.12 $ & $ 4.1 $ & $ 1.1 $ & $2.3 $ & $0.41 $ & $ 13 $\% ($ 1.5 \sigma$) $[2]$\\
\hline \textbf{expc3} & $ 0.36 $ & $ 14 $ & $ 0.81 $ &    --    & $ 0.69 $ & $ 0.97 $ & $ 3.2 $ & $ 0.12 $ & $ 4.5 $ & $ 1 $ & $3.6 $ & $-0.44 $ & $ 21 $\% ($ 1.2 \sigma$) $[3]$\\
\hline \textbf{ jump} & $ 0.37 $ &    --    & $ 0.81 $ &    --    & $ 0.69 $ & $ 0.97 $ & $ 3.2 $ & $ 0.12 $ & $ 4.6 $ & $ 1 $ & $2.9 $ & $0.3 $ & $  10 $\% ($ 1.6 \sigma$) $[2]$\\
\end{tabular}
\end{ruledtabular}
  \caption{\label{tab:TT_lowP_H0MP} The same as Table~\ref{tab:TT_onlyMP} but with Planck $TT$ + lowP + H0.}
\end{table*}


\begin{table*}
  \begin{ruledtabular}
  \begin{tabular}{ c | c | c | c | c | c | c | c | c | c | c | c | c | r }
  $\xi(k)$ & $k_\mathrm{c}^{-1} [\mathrm{Gpc}]$ & $\lambda$ & $\beta$ & $\alpha$ & $h$ & $\ns$ & $\ln(10^{10}\As)$ & $\tau$ & $\Gamma$ & $\Gamma_\mathrm{lowP}$ & $\Gamma_\mathrm{high}$ & $\Gamma_\mathrm{BAO}$ & $\gamma$ \\
\hline \textbf{ plaw} &    --    &    --    &    --    &    --    & $ 0.68 $ & $ 0.97 $ & $ 3.1 $ & $ 0.088 $ & $ 0 $ & $ 0 $ & $0 $ & $ 0 $ &        --      \\
\hline \textbf{  bpl} & $ 1.2 $ & $ 0.63 $ &    --    &    --    & $ 0.68 $ & $ 0.97 $ & $ 3.1 $ & $ 0.096 $ & $ 2.3 $ & $ 1.2 $ & $1.1 $ & $ 0.06 $ & $ 31 $\% ($ 1 \sigma$) $[2]$\\
\hline \textbf{atan1} & $ 5.3 $ & $ 1 $ & $ 0 $ & $ 1 $ & $ 0.68 $ & $ 0.96 $ & $ 3.2 $ & $ 0.11 $ & $ 3.5 $ & $ 1.8 $ & $1.7 $ & $ 0.08 $ & $  6 $\% ($ 1.9 \sigma$) $[1]$\\
\hline \textbf{atan2} & $ 5.3 $ & $ 1 $ & $ 0 $ & $ 1 $ & $ 0.68 $ & $ 0.96 $ & $ 3.2 $ & $ 0.11 $ & $ 3.5 $ & $ 1.8 $ & $1.7 $ & $ 0.08 $ & $ 17 $\% ($ 1.4 \sigma$) $[2]$\\
\hline \textbf{atan3} & $ 5.3 $ & $ 1 $ & $ 0 $ & $ 1 $ & $ 0.68 $ & $ 0.96 $ & $ 3.2 $ & $ 0.11 $ & $ 3.5 $ & $ 1.8 $ & $1.7 $ & $ 0.08 $ & $ 32 $\% ($ 1 \sigma$) $[3]$\\
\hline \textbf{atan4} & $ 0.37 $ & $ 46 $ & $ 0.82 $ & $ 0.52 $ & $ 0.68 $ & $ 0.97  $ & $ 3.2 $ & $ 0.11 $ & $ 4.6 $ & $ 2.4 $ & $1.9 $ & $ 0.14 $ & $ 33 $\% ($ 0.97\sigma$) $[4]$\\
\hline \textbf{expc1} & $ 3 $ & $ 0.5 $ & $ 0 $ &    --    & $ 0.68 $ & $ 0.97 $ & $ 3.2 $ & $ 0.11 $ & $ 3.8 $ & $ 1.6 $ & $2 $ & $ 0.05 $ & $  5 $\% ($ 2 \sigma$) $[1]$\\
\hline \textbf{expc2} & $ 3 $ & $ 0.53 $ & $ 0 $ &    --    & $ 0.68 $ & $ 0.97 $ & $ 3.1 $ & $ 0.11 $ & $ 3.9 $ & $ 2.1 $ & $1.6 $ & $ 0.08 $ & $ 14 $\% ($ 1.5 \sigma$) $[2]$\\
\hline \textbf{expc3} & $ 0.35 $ & $ 17 $ & $ 0.8 $ &    --    & $ 0.68 $ & $ 0.97 $ & $ 3.2 $ & $ 0.12 $ & $ 4.4 $ & $ 1.6 $ & $2.4 $ & $ 0.1 $ & $ 22 $\% ($ 1.2 \sigma$) $[3]$\\
\hline \textbf{ jump} & $ 0.37 $ &    --    & $ 0.8 $ &    --    & $ 0.68 $ & $ 0.97 $ & $ 3.2 $ & $ 0.12 $ & $ 4.3 $ & $ 1.4 $ & $2.6 $ & $ 0.11 $ & $ 11 $\% ($ 1.6 \sigma$) $[2]$\\
\end{tabular}
\end{ruledtabular}
  \caption{\label{tab:TT_lowP_BAOMP} The same as Table~\ref{tab:TT_onlyMP} but with Planck $TT$ + lowP + BAO.}
\end{table*}


\begin{table*}[ht!]
  \begin{ruledtabular}
  \begin{tabular}{ c | c | c | c | c | c | c | c | c | c | c | c | c | c | r }
  $\xi(k)$ & $k_\mathrm{c}^{-1} [\mathrm{Gpc}]$ & $\lambda$ & $\beta$ & $\alpha$ & $h$ & $\ns$ & $\ln(10^{10}\As)$ & $\tau$ & $\Gamma$ & $\Gamma_\mathrm{lowP}$ & $\Gamma_\mathrm{high}$ & $\Gamma_{H0}$ & $\Gamma_\mathrm{BAO}$ & $\gamma$ \\
\hline \textbf{ plaw} &    --    &    --    &    --    &    --    & $ 0.69 $ & $ 0.97 $ & $ 3.1 $ & $ 0.098 $ & $ 0 $ & $ 0 $ & $0 $ & $0 $ & $ 0 $ &        --      \\
\hline \textbf{  bpl} & $ 1.2 $ & $ 0.69 $ &    --    &    --    & $ 0.69 $ & $ 0.97 $ & $ 3.1 $ & $ 0.1 $ & $ 2.9 $ & $ 2.5 $ & $0.14 $ & $0.34 $ & $-0.12 $ & $ 24 $\% ($ 1.2 \sigma$) $[2]$\\
\hline \textbf{atan1} & $ 5.7 $ & $ 1 $ & $ 0 $ & $ 1 $ & $ 0.69 $ & $ 0.97 $ & $ 3.1 $ & $ 0.1 $ & $ 3.9 $ & $ 3.8 $ & $0.02 $ & $0.17 $ & $ 0.01 $ & $  4.9 $\% ($ 2 \sigma$) $[1]$\\
\hline \textbf{atan2} & $ 5.7 $ & $ 1 $ & $ 0 $ & $ 1 $ & $ 0.69 $ & $ 0.97 $ & $ 3.1 $ & $ 0.1 $ & $ 3.9 $ & $ 3.8 $ & $0.02 $ & $0.17 $ & $ 0.01 $ & $ 14 $\% ($ 1.5 \sigma$) $[2]$\\
\hline \textbf{atan3} & $ 5.7 $ & $ 1 $ & $ 0 $ & $ 1 $ & $ 0.69 $ & $ 0.97 $ & $ 3.1 $ & $ 0.1 $ & $ 3.9 $ & $ 3.8 $ & $0.03 $ & $0.17 $ & $ 0.01 $ & $ 27 $\% ($ 1.1 \sigma$) $[3]$\\
\hline \textbf{atan4} & $ 0.38 $ & $ 42 $ & $ 0.82 $ & $ 0.52 $ & $ 0.69 $ & $ 0.97 $ & $ 3.2 $ & $ 0.11 $ & $ 5 $ & $ 3.5 $ & $1.1 $ & $0.43 $ & $-0.15 $ & $ 29 $\% ($ 1.1 \sigma$) $[4]$\\
\hline \textbf{expc1} & $ 3.1 $ & $ 0.5 $ & $ 0 $ &    --    & $ 0.69 $ & $ 0.97 $ & $ 3.2 $ & $ 0.11 $ & $ 4.2 $ & $ 2.9 $ & $1.1 $ & $0.17 $ & $ 0 $ & $  4 $\% ($ 2 \sigma$) $[1]$\\
\hline \textbf{expc2} & $ 3.1 $ & $ 0.5 $ & $ 0 $ &    --    & $ 0.69 $ & $ 0.97 $ & $ 3.2 $ & $ 0.11 $ & $ 4.2 $ & $ 2.9 $ & $1.1 $ & $0.17 $ & $ 0 $ & $ 12 $\% ($ 1.6 \sigma$) $[2]$\\
\hline \textbf{expc3} & $ 0.36 $ & $ 16 $ & $ 0.81 $ &    --    & $ 0.69 $ & $ 0.97 $ & $ 3.2 $ & $ 0.12 $ & $ 4.7 $ & $ 2.6 $ & $1.7 $ & $0.3 $ & $-0.05 $ & $ 19 $\% ($ 1.3 \sigma$) $[3]$\\
\hline \textbf{ jump} & $ 0.37 $ &    --    & $ 0.81 $ &    --    & $ 0.69 $ & $ 0.97 $ & $ 3.2 $ & $ 0.12 $ & $ 4.8 $ & $ 2.5 $ & $1.8 $ & $0.3 $ & $-0.04 $ & $  9.2 $\% ($ 1.7 \sigma$) $[2]$\\
\end{tabular}
\end{ruledtabular}
  \caption{\label{tab:TT_lowP_H0_BAOMP} The same as Table~\ref{tab:TT_onlyMP} but with Planck $TT$ + lowP + H0 + BAO.}
\end{table*}

The standard Bayesian approach in this case uses the
Bayes factor
\begin{equation}
B_\textsc{fa} = \frac{\displaystyle\int\dd\theta\;
L(D|\hat\theta)P(\theta)}{\displaystyle\int\dd\theta_\mathrm{F}\;
L(D|\hat\theta_\mathrm{F})P_\mathrm{F}(\theta_\mathrm{F})},
\end{equation}
where $P_\mathrm{F}$ and $P$ are the respective priors in the fiducial
and alternative models. As in Ref.~\cite{Ade:2015lrj}, if we
consider the same flat prior for both models
($P_\mathrm{F}(\theta_\mathrm{F}) = 1$ and $P(\theta) = 1$, besides the
PFI priors cited above), the LRT gives us a point estimate of the Bayes
factor:
\begin{equation*}
B_\textsc{fa} \approx \exp\left(\frac{\Gamma}{2}\right).
\end{equation*}

In this work we do not initially follow the Bayesian approach for all
the models for two reasons. First, as we stated above, in this
phenomenological study we want to understand the ability of the current
data to inform us about different aspects of the model, whereas a
Bayesian approach would simply tend to penalize any irrelevant extra
parameters in the alternative model. Our initial interest is in
determining if those extra parameters should be included in the
analysis. The second reason why we initially perform a frequentist
analysis stems, again, from the phenomenological nature of our approach:
we may not have any theoretical reason to assume a specific prior for
the extra parameters, and in fact even a flat prior would not be
unambiguous since it depends on the choice of parameters. Note that the
frequentist approach does not depend on the introduction of a measure in
the model space (usually done through simple priors in the model
parameters).

A frequentist approach does not answer the same questions as a Bayesian
methodology. All we can know in a frequentist study is the ability of
the current data to falsify the fiducial model (the null hypothesis). In
other words, if an alternative model provides a better fit to the
data$-$one that goes beyond the improvement expected from statistical
fluctuations and from the addition of extra parameters$-$then we may say
that the fiducial model is falsified.\footnote{Of course, to obtain the
relevant probabilities it is necessary to simulate a large number of
samples from the fiducial and alternative models, or to use Wilks'
theorem which only includes the probability of the data under the null
hypothesis.} Since we will be evaluating a large number of cases, we
chose to first follow the frequentist approach, thereby answering the
simplest questions while avoiding the introduction of a measure in the
model space. We then apply a follow-up Bayesian analysis to what appears
to be the best competing model.

In our Bayesian approach we run a complete MCMC analysis of the posteriors of
the fiducial and competing models using an ensemble sampler algorithm that was
introduced in Ref.~\cite{Goodman2010}\footnote{There also exists a P\textsc{ython}
	implementation, see Ref.~\cite{Foreman-Mackey2013}.} and which is here
implemented in NumCosmo as described in Appendix~\ref{app:numcosmo}. From the
results we produce a corner plot containing the marginal distributions and
two-dimensional confidence regions for all relevant parameters. We then apply
the modified harmonic mean, also described in Appendix~\ref{app:numcosmo}, to
estimate the Bayes factor resulting from the comparison of the fiducial and
competing models.

\begin{figure}[ht]
	\begin{center}
		\includegraphics[scale=1.0]{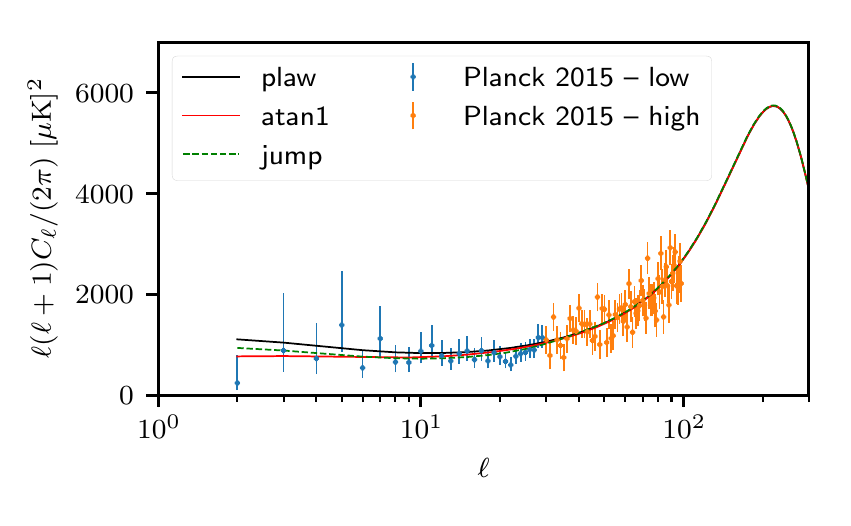}
	\end{center}
	\caption{Examples of best-fit CMB anisotropies with the different deficit
	functions (see text), together with the Planck 2015 $TT$ data up to $\ell =
	100$ (this limit is chosen for aesthetic reasons). The smooth transition
	represented by \textbf{atan1} results in a larger deficit at the lower
	multipoles ($\ell = $2--10), whereas \textbf{jump} results in a deficit which
	is smaller at these lower multipoles and larger at the higher multipoles
	$\ell=$20--30. This reveals the difference between the sharp and the smooth
	deficit. The latter improves the fit at the lowest multipoles only, while the
	former also affects the spectrum around $\ell=$20--30 where the cosmic variance
	is less important. These fits are for the full Planck dataset. \label{fig:Cls}}
\end{figure}

We emphasize that the main objective of this paper is to compute
the value of $\Gamma$ for each alternative model (and
$\Gamma_i$ for each component of the final likelihood). The value of $B_\textsc{fa}$ is computed only for the \textbf{jump} model. Naturally, we also obtain the best-fit parameters and (from the MCMC analysis) the full posterior. While a parameter-space analysis can be useful to understand the behavior of a model, this is not our main objective.

\begin{figure*}
	\begin{center}
		\includegraphics[scale=0.5]{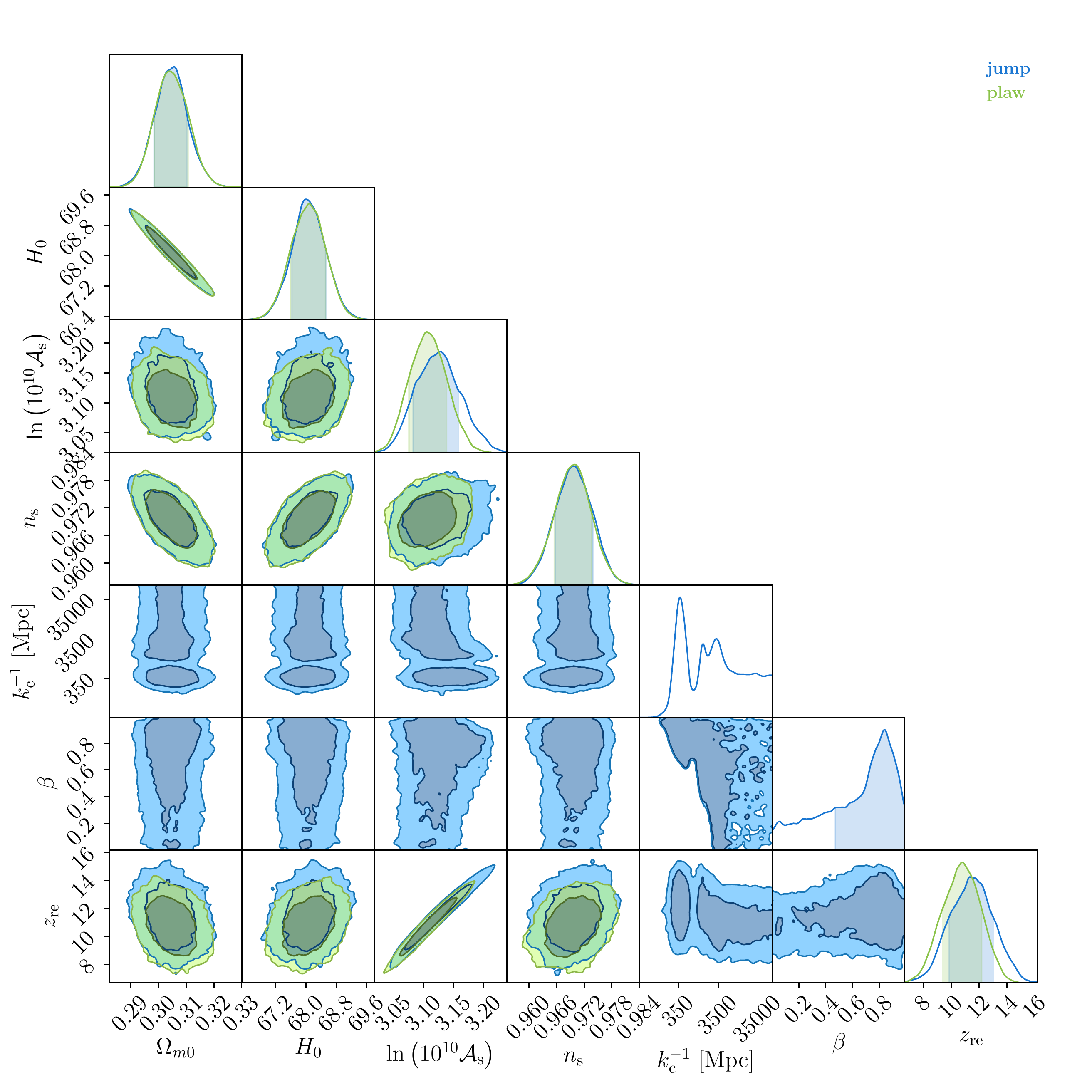} 
	\end{center}
	\caption{Corner plot for the MCMC results with \textbf{plaw} and \textbf{jump}
	using the full Planck dataset, H0 and BAO.  The marginal distribution for
	$k_\mathrm{c}^{-1}$ has three modes: the first at $\sim 380$~Mpc, the second
	at $\sim 1450$~Mpc and the third at $\sim 3200$~Mpc. As expected the
	distribution for $k_\mathrm{c}^{-1}$ becomes flat for $k_\mathrm{c}^{-1}
	\gtrsim 3200 \mathrm{Mpc}$, since at these values the \textbf{jump} mode is
	numerically equivalent to \textbf{plaw}. Furthermore, in this interval, the
	distribution for $\beta$ becomes completely degenerate as can be seen in the
	$k_\mathrm{c}^{-1} \times \beta$ confidence region. This is the reason why the
	short-distance mode appears to provide a definite set of values and may
	therefore be considered physically relevant, while the second and third (long-distance) modes are too degenerate in the actual value of the amplitude of the jump (not to mention that the scale induced is very close to the Hubble scale). Thus we do not consider these modes any further. \label{corner:full}}
\end{figure*}

\section{Numerical approach} \label{sec:num}

The theoretical CMB anisotropies $C^{XY}_\ell$ do not depend on the Planck
foreground and instrument parameters in $\theta_\mathrm{PFI}$. For this reason
we divide the problem of finding the best fit into two steps. In the first step,
we define the full Planck likelihood $L(D | \theta)$ as a function on the whole
parameter space. Fixing the values of $\{ \theta_{\Lambda\mathrm{CDM}},
\theta_{\xi} \}$, we can cheaply calculate the likelihood for different points
of $\theta_\mathrm{PFI}$ since we can re-use the same $C^{XY}_\ell$. We then
define the PFI-Likelihood as
\begin{equation}\label{LPFI}
L_\mathrm{PFI}\left(D|\theta_{\Lambda\mathrm{CDM}},\theta_{\xi}\right) =
\max_{\theta_\mathrm{PFI}}L(D|\theta).
\end{equation}
In the second step, we find the maximum of $ L_\mathrm{PFI}
\left(D|\theta_{\Lambda\mathrm{CDM}}, \theta_{\xi}\right) $. Since these
two steps are mathematically equivalent (when the likelihood is
$C^\infty$ in the parameters), we can use them to speed up significantly
the finding of the best fit. Furthermore, in a multimodal likelihood
(which is frequently the case for high-dimensional likelihoods) the
fitting process can always stop prematurely for two reasons. First, if
it finds a local maximum. Second, if it is moving in a plateau where the
likelihood varies slowly (that is, where it has a very small gradient).
There is no known algorithm that could guarantee that the true maximum
has been found. Thus we have applied the usual checking method of
starting the fitting process from different initial points in the
parametric space in order to minimize this risk. The specific objects
and methods used at this stage are described in
Appendix~\ref{app:numcosmo}.

\begin{figure*}
	\begin{center}
		\includegraphics[scale=0.5]{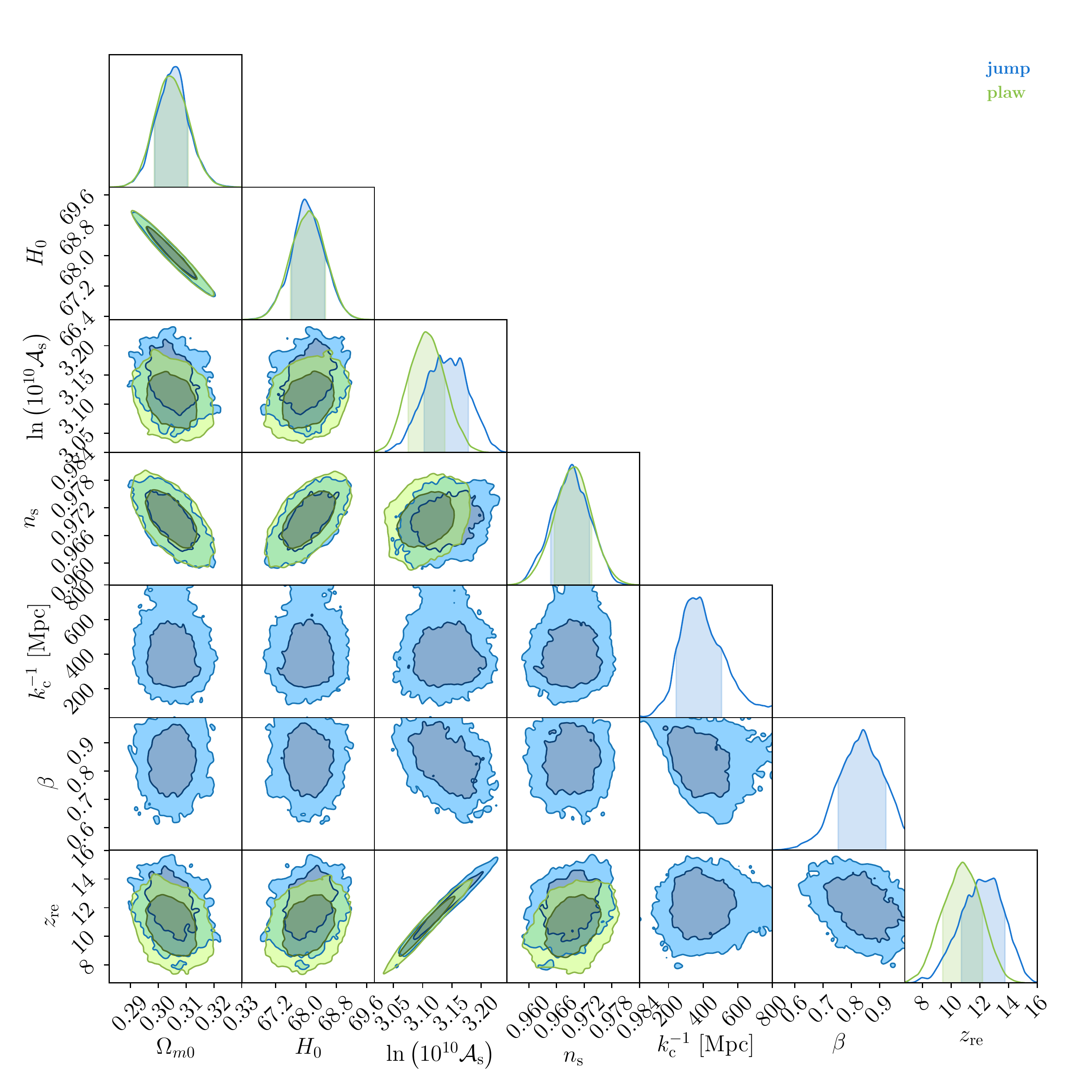} 
	\end{center}
	\caption{Corner plot for the MCMC results from \textbf{plaw} and \textbf{jump}
	using the full Planck dataset, H0 and BAO and removing the second and third modes
	(see~Fig.~\ref{corner:full}). In this plot we use a linear scale for
	$k_\mathrm{c}^{-1}$. Note that once the second mode is removed, there is a
	shift in $\mathcal{A}_\mathrm{s}$ and $z_\mathrm{re}$. This was already present
	in Fig~\ref{corner:full}, in the $k_\mathrm{c}^{-1} \times
	\log(10^{10}\mathcal{A}_\mathrm{s})$ and $k_\mathrm{c}^{-1} \times
	z_\mathrm{re}$ confidence regions, where the one-sigma contour related to the
	second mode of $k_\mathrm{c}^{-1}$ is slightly shifted when compared to the
	first mode. In addition, we have a negative correlation for $k_\mathrm{c}^{-1}
	\times \beta$: a smaller jump scale $k_\mathrm{c}^{-1}$ implies a smaller jump
	amplitude ($\beta$ closer to unity). \label{corner} }
\end{figure*}

For the computation of $C^{XY}_\ell$ we employed the C\textsc{lass} back
end of NumCosmo (see
Refs.~\cite{Blas2011,Lesgourgues2011a,Lesgourgues2011b}). The precision
settings were increased compared to the default configuration; further
details can be found in Appendix~\ref{app:numcosmo}. It suffices to note
here that, in order to measure differences between primordial power
spectra, we considered three different precision settings: low precision
(LP) (equal to the default setting in CLASS, as in their version 2.5.0),
medium precision (MP), and high precision (HP). Increasing the precision
from low to medium changes the results in some cases (with $\Delta\Gamma
\approx 0.5$), but the results then remain stable when the precision is
further increased to the highest level. We therefore report results
obtained with medium precision.

\begin{figure}[ht]
	\begin{center}
		\includegraphics[scale=1]{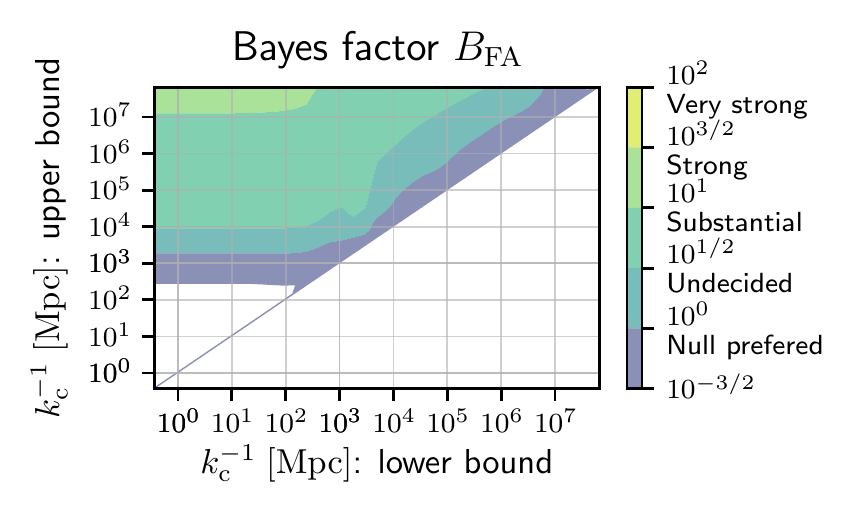}
	\end{center}	
	\caption{Bayesian evidence comparing \textbf{jump} and \textbf{plaw} models for
	various combinations of upper and lower bounds on the $k_\mathrm{c}^{-1}$
	prior. This analysis used the full Planck dataset, H0 and BAO-derived distances. The white areas represent forbidden regions (lower bound greater than upper bound) or regions without enough points in the sample. To the right of the graph we include the evidence scale (Jeffreys' scale) proposed in	Ref.~\cite[Appendix B]{Jeffreys1998}.\label{Bfactor}}
\end{figure}

\section{Results} \label{sec:results}

We group our results according to which CMB data we are using. For a
given CMB sample we discuss its results alone and in combination with
the other samples H0, BAO and H0+BAO. We already know from previous
studies (see for example~\cite{Ade:2015lrj} and references therein)
that there is a lack of power on large scales for the temperature data. However, studying this effect alone can be misleading as it is difficult to take into account the look-elsewhere effect when we are dealing with a large and heterogeneous body of data. We therefore compare the whole fit when using different datasets, and we also include the differences in the fit for each relevant part of the likelihood. This analysis of the significance of the different deficit models suggests that most of them are not sufficiently competitive to justify a complete MCMC analysis of their parametric space.

\subsection{General considerations}

We summarize our results in Tables \ref{tab:TT_onlyMP} to
\ref{tab:TTTEEE_lowP_H0_BAOMP}. For each parametrization and dataset used, we
show the best-fit values of the cutoff scale $k_\mathrm{c}^{-1}$ in Gpc and
(when relevant) the values of the dimensionless parameters $\lambda$, $\beta$
and $\alpha$. The first ($\lambda$) measures the sharpness of the deficit
function, the second ($\beta$) quantifies the deficit, while the third
($\alpha$) describes the shape of the transition. After these four new
parameters we include the best-fit values of the four standard parameters: the
dimensionless rescaled Hubble scale $h$, the spectral index $n_\mathrm{s}$ and
amplitude $A_\mathrm{s}$ of the fiducial PPS \eqref{eq:PPS:fiduc}, and finally
the reionization optical depth $\tau$ (which is computed as a derived quantity
from the best-fit values). To complete the tables we add the LRT statistics
$\Gamma$ defined by \eqref{LRTS} and (in the last column) the $p-$value $\gamma$
defined by \eqref{gamma}. To the last column we add (in round brackets) the
probability $\gamma$ translated into the corresponding number of one-dimensional
Gaussian standard deviations, and also (in square brackets) the number of extra
degrees of freedom. It is often assumed that a $p-$value of 5\% indicates
statistical significance. We shall here instead take the view that whenever a
small value is found, this merely suggests a plausible new effect.

In the first line of Table~\ref{tab:TTTEEE_lowPMP} we reproduce the
currently accepted values for the four standard-model parameters $h$,
$n_\mathrm{s}$, $A_\mathrm{s}$ and $\tau$, where our fit matches that
provided by the Planck team (see, however, App~\ref{app:numcosmo}).

The first conclusion one can draw from the tables is that the standard-model
parameters are hardly affected by the inclusion of the deficit function,
regardless of the choice of the latter. This shows that even if the lack of
power is a true physical phenomenon, it cannot come from a strong effect as
otherwise the analysis would have shown some instability when including this
phenomenon in the description of the data. This also immediately shows that
there is no chance of resolving the $H_0$ tension by taking the deficit function
into account. Such a possibility might be considered upon examining
Table~\ref{tab:TT_only_H0MP}, where we see that a reasonable amount of the
significance is produced by a better fitting of the H0 data (see~$\Gamma_{H0}$).
However, this effect is severely reduced when polarization data are added. In
other words, when fitting $TT$ data alone the extra freedom in the PPS seems to
allow a better fit of the $TT$ + H0 data combination, but this does not hold
when polarization data are added. On the contrary, comparing with the
\textbf{plaw} fits including H0, we see that when polarization data (at both
low-$\ell$ and high-$\ell$) are added we get $-2\ln(L_{H0}) \approx 9$, which
corresponds roughly to the well-known $3\sigma$ tension with the H0 data. For
this reason, our results including H0 should be interpreted with caution.

\subsection{The smooth deficit functions}

Throughout the data, if we set the interval for the parameter $\lambda$ to be
around unity\footnote{This is not to be considered a prior as it is only a
	feature of the minimization algorithm. If in any event the minimization process
	takes the best-fit close to the boundaries of these intervals, then they should
	be extended and the minimization rerun.} (that is, if we impose a smooth
transition), it is found that the best-fit value of the transition scale remains
very close to the Hubble radius (in terms of $k^{-1}$, corresponding to a length scale one order of magnitude larger than the Hubble radius) -- where the data are dominated by cosmic variance. As a result, these fits
are only marginally significant (like those originally discussed by the Planck
team). Adding data for polarization, H0 and the BAO's does not change this trend,
and in fact adding polarization actually reduces the significance.

Our weakly-significant fits neither support nor rule out the smooth deficit
functions we have studied. Statistically speaking, according to our frequentist
analysis, these functions are more or less as successful as the standard
power-law model (taking into account the larger number of parameters). This
result might be viewed as a modest success, in the sense that a smooth deficit
could have been disfavored compared to no deficit but instead performs
comparably well. On the other hand, the data we have studied are consistent with
the apparent low power being a mere statistical fluctuation. In the absence of a
significantly better fit for the models with a smooth deficit, it is also
natural to invoke Ockham's razor to favor the simpler model with no extra
parameters (even if, strictly speaking, no such conclusion can be drawn on the
basis of the significance of the fits).

Furthermore, the data we have studied certainly cannot constrain the shape of
the smooth-deficit spectrum even if it exists. This seems to be so regardless of
the functional form of the deficit function ({\bf expc}, {\bf bpl} or {\bf
	atan}), and independently of the number of extra parameters and priors assumed.
This conclusion is compatible with our result that the shape parameter $\alpha$
is essentially irrelevant. Roughly speaking, we found $\alpha$ ranging between
$\frac12$ and 1. As shown in Fig.~\ref{fig:xi_cmp}, this amounts to hardly any
variation at all in the actual spectrum. Thus it appears that the data cannot
favor any particular shape.


\begin{table*}
  \begin{ruledtabular}
  \begin{tabular}{ c | c | c | c | c | c | c | c | c | c | c | c | r }
  $\xi(k)$ & $k_\mathrm{c}^{-1} [\mathrm{Gpc}]$ & $\lambda$ & $\beta$ & $\alpha$ & $h$ & $\ns$ & $\ln(10^{10}\As)$ & $\tau$ & $\Gamma$ & $\Gamma_\mathrm{lowP}$ & $\Gamma_\mathrm{highP}$ & $\gamma$ \\
\hline \textbf{ plaw} &    --    &    --    &    --    &    --    & $ 0.67 $ & $ 0.97 $ & $ 3.1 $ & $ 0.082 $ & $ 0 $ & $ 0 $ & $0 $ &        --      \\
\hline \textbf{  bpl} & $ 0.21 $ & $ 0.14 $ &    --    &    --    & $ 0.67 $ & $ 0.96 $ & $ 3.1 $ & $ 0.1 $ & $ 4.7 $ & $ 2.6 $ & $1.9 $ & $  9.6 $\% ($ 1.7 \sigma$) $[2]$\\
\hline \textbf{atan1} & $ 6.5 $ & $ 1 $ & $ 0 $ & $ 1 $ & $ 0.67 $ & $ 0.96 $ & $ 3.1 $ & $ 0.094 $ & $ 3.4 $ & $ 2.8 $ & $0.65 $ & $  6.7 $\% ($ 1.8 \sigma$) $[1]$\\
\hline \textbf{atan2} & $ 6.5 $ & $ 1 $ & $ 0 $ & $ 1 $ & $ 0.67 $ & $ 0.96 $ & $ 3.1 $ & $ 0.094 $ & $ 3.4 $ & $ 2.8 $ & $0.82 $ & $ 19 $\% ($ 1.3 \sigma$) $[2]$\\
\hline \textbf{atan3} & $ 6.5 $ & $ 1 $ & $ 0.003 $ & $ 0.98 $ & $ 0.67 $ & $ 0.96  $ & $ 3.1 $ & $ 0.094 $ & $ 3.4 $ & $ 2.8 $ & $0.76 $ & $ 34 $\% ($ 0.96\sigma$) $[3]$\\
\hline \textbf{atan4} & $ 0.38 $ & $ 31 $ & $ 0.82 $ & $ 0.7 $ & $ 0.67 $ & $ 0.96 $ & $ 3.1 $ & $ 0.091 $ & $ 5.7 $ & $ 4.3 $ & $1.3 $ & $ 22 $\% ($ 1.2 \sigma$) $[4]$\\
\hline \textbf{expc1} & $ 3.4 $ & $ 0.5 $ & $ 0 $ &    --    & $ 0.67 $ & $ 0.96 $ & $ 3.1 $ & $ 0.1 $ & $ 4 $ & $ 2.4 $ & $1.6 $ & $  4.5 $\% ($ 2 \sigma$) $[1]$\\
\hline \textbf{expc2} & $ 2.7 $ & $ 0.57 $ & $ 0 $ &    --    & $ 0.67 $ & $ 0.96 $ & $ 3.1 $ & $ 0.099 $ & $ 4.2 $ & $ 2.3 $ & $1.8 $ & $ 13 $\% ($ 1.5 \sigma$) $[2]$\\
\hline \textbf{expc3} & $ 0.35 $ & $ 13 $ & $ 0.82 $ &    --    & $ 0.67 $ & $ 0.96 $ & $ 3.1 $ & $ 0.096 $ & $ 5.7 $ & $ 3.8 $ & $1.7 $ & $ 13 $\% ($ 1.5 \sigma$) $[3]$\\
\hline \textbf{ jump} & $ 0.38 $ &    --    & $ 0.82 $ &    --    & $ 0.67 $ & $ 0.96 $ & $ 3.1 $ & $ 0.093 $ & $ 5.8 $ & $ 4.2 $ & $1.5 $ & $  5.5 $\% ($ 1.9 \sigma$) $[2]$\\
\end{tabular}
\end{ruledtabular}
  \caption{\label{tab:TTTEEE_lowPMP} The same as Table~\ref{tab:TT_onlyMP} but with Planck $TT$, $TE$, $EE$ + lowP.}
\end{table*}


\begin{table*}
  \begin{ruledtabular}
  \begin{tabular}{ c | c | c | c | c | c | c | c | c | c | c | c | c | r }
  $\xi(k)$ & $k_\mathrm{c}^{-1} [\mathrm{Gpc}]$ & $\lambda$ & $\beta$ & $\alpha$ & $h$ & $\ns$ & $\ln(10^{10}\As)$ & $\tau$ & $\Gamma$ & $\Gamma_\mathrm{lowP}$ & $\Gamma_\mathrm{highP}$ & $\Gamma_{H0}$ & $\gamma$ \\
\hline \textbf{ plaw} &    --    &    --    &    --    &    --    & $ 0.68 $ & $ 0.97 $ & $ 3.1 $ & $ 0.089 $ & $ 0 $ & $ 0 $ & $0 $ & $0 $ &        --      \\
\hline \textbf{  bpl} & $ 0.22 $ & $ 0.14 $ &    --    &    --    & $ 0.68 $ & $ 0.97 $ & $ 3.1 $ & $ 0.1 $ & $ 4.8 $ & $ 2.5 $ & $2.1 $ & $-0.01 $ & $  9.2 $\% ($ 1.7 \sigma$) $[2]$\\
\hline \textbf{atan1} & $ 7.3 $ & $ 1 $ & $ 0 $ & $ 1 $ & $ 0.68 $ & $ 0.96 $ & $ 3.1 $ & $ 0.096 $ & $ 3.2 $ & $ 3 $ & $0.33 $ & $0.01 $ & $  7.4 $\% ($ 1.8 \sigma$) $[1]$\\
\hline \textbf{atan2} & $ 6.8 $ & $ 1 $ & $ 0.022 $ & $ 1 $ & $ 0.68 $ & $ 0.96 $ & $ 3.1 $ & $ 0.098 $ & $ 3.2 $ & $ 2.9 $ & $0.66 $ & $-0.11 $ & $ 20 $\% ($ 1.3 \sigma$) $[2]$\\
\hline \textbf{atan3} & $ 4.8 $ & $ 1 $ & $ 0.005 $ & $ 0.51 $ & $ 0.68 $ & $ 0.96 $ & $ 3.1 $ & $ 0.094 $ & $ 3.3 $ & $ 3.2 $ & $0.23 $ & $0.02 $ & $ 35 $\% ($ 0.93 \sigma$) $[3]$\\
\hline \textbf{atan4} & $ 0.39 $ & $ 31 $ & $ 0.82 $ & $ 0.77 $ & $ 0.68 $ & $ 0.97 $ & $ 3.1 $ & $ 0.099 $ & $ 5.5 $ & $ 3.7 $ & $1.6 $ & $-0.06 $ & $ 24 $\% ($ 1.2 \sigma$) $[4]$\\
\hline \textbf{expc1} & $ 3.7 $ & $ 0.5 $ & $ 0 $ &    --    & $ 0.68 $ & $ 0.97 $ & $ 3.1 $ & $ 0.1 $ & $ 4.1 $ & $ 2.4 $ & $1.6 $ & $0.08 $ & $  4.2 $\% ($ 2 \sigma$) $[1]$\\
\hline \textbf{expc2} & $ 2.8 $ & $ 0.58 $ & $ 0 $ &    --    & $ 0.68 $ & $ 0.97 $ & $ 3.1 $ & $ 0.1 $ & $ 4.2 $ & $ 2.3 $ & $1.6 $ & $0.2 $ & $ 12 $\% ($ 1.5 \sigma$) $[2]$\\
\hline \textbf{expc3} & $ 0.36 $ & $ 14 $ & $ 0.82 $ &    --    & $ 0.68 $ & $ 0.97 $ & $ 3.1 $ & $ 0.1 $ & $ 5.5 $ & $ 3.3 $ & $2.2 $ & $-0.32 $ & $ 14 $\% ($ 1.5 \sigma$) $[3]$\\
\hline \textbf{ jump} & $ 0.39 $ &    --    & $ 0.82 $ &    --    & $ 0.68 $ & $ 0.97 $ & $ 3.1 $ & $ 0.095 $ & $ 5.3 $ & $ 4.1 $ & $0.99 $ & $0.08 $ & $  7.1 $\% ($ 1.8 \sigma$) $[2]$\\
\end{tabular}
\end{ruledtabular}
  \caption{\label{tab:TTTEEE_lowP_H0MP} The same as Table~\ref{tab:TT_onlyMP} but with Planck $TT$, $TE$, $EE$ + lowP + H0.}
\end{table*}


\begin{table*}
  \begin{ruledtabular}
  \begin{tabular}{ c | c | c | c | c | c | c | c | c | c | c | c | c | r }
  $\xi(k)$ & $k_\mathrm{c}^{-1} [\mathrm{Gpc}]$ & $\lambda$ & $\beta$ & $\alpha$ & $h$ & $\ns$ & $\ln(10^{10}\As)$ & $\tau$ & $\Gamma$ & $\Gamma_\mathrm{lowP}$ & $\Gamma_\mathrm{highP}$ & $\Gamma_\mathrm{BAO}$ & $\gamma$ \\
\hline \textbf{ plaw} &    --    &    --    &    --    &    --    & $ 0.68 $ & $ 0.97 $ & $ 3.1 $ & $ 0.085 $ & $ 0 $ & $ 0 $ & $0 $ & $ 0 $ &        --      \\
\hline \textbf{  bpl} & $ 0.23 $ & $ 0.14 $ &    --    &    --    & $ 0.68 $ & $ 0.97 $ & $ 3.1 $ & $ 0.1 $ & $ 4.7 $ & $ 2.3 $ & $2.2 $ & $-0.03 $ & $  9.4 $\% ($ 1.7 \sigma$) $[2]$\\
\hline \textbf{atan1} & $ 7 $ & $ 1 $ & $ 0 $ & $ 1 $ & $ 0.68 $ & $ 0.96 $ & $ 3.1 $ & $ 0.095 $ & $ 3.2 $ & $ 2.9 $ & $0.59 $ & $-0.08 $ & $  7.4 $\% ($ 1.8 \sigma$) $[1]$\\
\hline \textbf{atan2} & $ 7 $ & $ 1 $ & $ 0.013 $ & $ 1 $ & $ 0.68 $ & $ 0.96 $ & $ 3.1 $ & $ 0.097 $ & $ 3.2 $ & $ 2.7 $ & $0.73 $ & $-0.15 $ & $ 20 $\% ($ 1.3 \sigma$) $[2]$\\
\hline \textbf{atan3} & $ 4.4 $ & $ 1 $ & $ 0.002 $ & $ 0.51 $ & $ 0.68 $ & $ 0.96 $ & $ 3.1 $ & $ 0.094 $ & $ 3.4 $ & $ 3 $ & $0.86 $ & $-0.25 $ & $ 33 $\% ($ 0.97 \sigma$) $[3]$\\
\hline \textbf{atan4} & $ 0.37 $ & $ 32 $ & $ 0.82 $ & $ 0.68 $ & $ 0.68 $ & $ 0.97 $ & $ 3.1 $ & $ 0.1 $ & $ 5.5 $ & $ 3.5 $ & $1.9 $ & $-0.06 $ & $ 24 $\% ($ 1.2 \sigma$) $[4]$\\
\hline \textbf{expc1} & $ 3.5 $ & $ 0.5 $ & $ 0 $ &    --    & $ 0.68 $ & $ 0.97 $ & $ 3.1 $ & $ 0.1 $ & $ 4.1 $ & $ 2.4 $ & $2 $ & $-0.28 $ & $  4.3 $\% ($ 2 \sigma$) $[1]$\\
\hline \textbf{expc2} & $ 2.8 $ & $ 0.59 $ & $ 0 $ &    --    & $ 0.68 $ & $ 0.97 $ & $ 3.1 $ & $ 0.1 $ & $ 4.1 $ & $ 2.2 $ & $1.8 $ & $ 0.06 $ & $ 13 $\% ($ 1.5 \sigma$) $[2]$\\
\hline \textbf{expc3} & $ 0.35 $ & $ 14 $ & $ 0.81 $ &    --    & $ 0.68 $ & $ 0.97 $ & $ 3.1 $ & $ 0.1 $ & $ 5.4 $ & $ 3.4 $ & $1.8 $ & $ 0.09 $ & $ 15 $\% ($ 1.5 \sigma$) $[3]$\\
\hline \textbf{ jump} & $ 0.38 $ &    --    & $ 0.83 $ &    --    & $ 0.68 $ & $ 0.97 $ & $ 3.1 $ & $ 0.093 $ & $ 5.3 $ & $ 4.2 $ & $1.1 $ & $-0.07 $ & $  7.1 $\% ($ 1.8 \sigma$) $[2]$\\
\end{tabular}
\end{ruledtabular}
  \caption{\label{tab:TTTEEE_lowP_BAOMP} The same as Table~\ref{tab:TT_onlyMP} but with Planck $TT$, $TE$, $EE$ + lowP + BAO.}
\end{table*}


\begin{table*}
  \begin{ruledtabular}
  \begin{tabular}{ c | c | c | c | c | c | c | c | c | c | c | c | c | c | r }
  $\xi(k)$ & $k_\mathrm{c}^{-1} [\mathrm{Gpc}]$ & $\lambda$ & $\beta$ & $\alpha$ & $h$ & $\ns$ & $\ln(10^{10}\As)$ & $\tau$ & $\Gamma$ & $\Gamma_\mathrm{lowP}$ & $\Gamma_\mathrm{highP}$ & $\Gamma_{H0}$ & $\Gamma_\mathrm{BAO}$ & $\gamma$ \\
\hline \textbf{ plaw} &    --    &    --    &    --    &    --    & $ 0.68 $ & $ 0.97 $ & $ 3.1 $ & $ 0.095 $ & $ 0 $ & $ 0 $ & $0 $ & $0 $ & $ 0 $ &        --      \\
\hline \textbf{  bpl} & $ 0.23 $ & $ 0.15 $ &    --    &    --    & $ 0.68 $ & $ 0.97 $ & $ 3.2 $ & $ 0.11  $ & $ 4.7 $ & $ 2.7 $ & $1.7 $ & $0.11$ & $ 0.01 $ & $  9.7 $\% ($ 1.7 \sigma$) $[2]$\\
\hline \textbf{atan1} & $ 6.6 $ & $ 1 $ & $ 0 $ & $ 1 $ & $ 0.68 $ & $ 0.96 $ & $ 3.1 $ & $ 0.099 $ & $ 3.2 $ & $ 4 $ & $-0.28 $ & $-0.16 $ & $-0.04 $ & $  7.1 $\% ($ 1.8 \sigma$) $[1]$\\
\hline \textbf{atan2} & $ 6.6 $ & $ 1 $ & $ 0 $ & $ 1 $ & $ 0.68 $ & $ 0.96 $ & $ 3.1 $ & $ 0.099 $ & $ 3.3 $ & $ 4 $ & $-0.26 $ & $-0.16 $ & $-0.04 $ & $ 20 $\% ($ 1.3 \sigma$) $[2]$\\
\hline \textbf{atan3} & $ 6.6 $ & $ 1 $ & $ 0 $ & $ 1 $ & $ 0.68 $ & $ 0.96 $ & $ 3.1 $ & $ 0.099 $ & $ 3.3 $ & $ 4 $ & $-0.24 $ & $-0.16 $ & $-0.04 $ & $ 35 $\% ($ 0.93 \sigma$) $[3]$\\
\hline \textbf{atan4} & $ 0.39 $ & $ 32 $ & $ 0.82 $ & $ 0.76 $ & $ 0.68 $ & $ 0.97 $ & $ 3.1 $ & $ 0.1 $ & $ 5.4 $ & $ 4.6 $ & $1.3 $ & $-0.4 $ & $-0.09 $ & $ 25 $\% ($ 1.2 \sigma$) $[4]$\\
\hline \textbf{expc1} & $ 3.7 $ & $ 0.5 $ & $ 0 $ &    --    & $ 0.68 $ & $ 0.97 $ & $ 3.1 $ & $ 0.1 $ & $ 4 $ & $ 3.5 $ & $0.51 $ & $0.02 $ & $ 0 $ & $  4.6 $\% ($ 2 \sigma$) $[1]$\\
\hline \textbf{expc2} & $ 2.6 $ & $ 0.6 $ & $ 0 $ &    --    & $ 0.68 $ & $ 0.97 $ & $ 3.1 $ & $ 0.11 $ & $ 4.3 $ & $ 2.9 $ & $1.2 $ & $0.12 $ & $ 0.01 $ & $ 12 $\% ($ 1.6 \sigma$) $[2]$\\
\hline \textbf{expc3} & $ 0.36 $ & $ 14 $ & $ 0.81 $ &    --    & $ 0.68 $ & $ 0.97 $ & $ 3.1 $ & $ 0.1 $ & $ 5.3 $ & $ 4.5 $ & $1.4 $ & $-0.5 $ & $-0.13 $ & $ 15 $\% ($ 1.4 \sigma$) $[3]$\\
\hline \textbf{ jump} & $ 0.38 $ &    --    & $ 0.83 $ &    --    & $ 0.68 $ & $ 0.97 $ & $ 3.1 $ & $ 0.095 $ & $ 5.1 $ & $ 5.1 $ & $0.23 $ & $-0.17 $ & $-0.02 $ & $  7.7 $\% ($ 1.8 \sigma$) $[2]$\\
\end{tabular}
\end{ruledtabular}
  \caption{\label{tab:TTTEEE_lowP_H0_BAOMP} The same as Table~\ref{tab:TT_onlyMP} but with Planck $TT$, $TE$, $EE$ + lowP + H0 + BAO.}
\end{table*}

\subsection{The jump function} \label{sec:jump}

Having found that a smooth transition to large-scale low power does not improve
the fit to a degree that is convincingly significant, we now consider the
extreme case of a sharp transition. More precisely, we discuss a transition that
is so fast that the data are incapable of discerning its structure.
Statistically speaking one could argue, from the results in our Tables, that the
jump case is not necessarily preferable to some of our other models such as {\bf
	atan1} or {\bf expc1}. However, one should bear in mind that the latter cases
correspond to imposing strong (delta function) priors on the extra parameters
and with no particular physical justification (though see Sec.~\ref{sec:QR}
for a possible physical motivation for {\bf atan1}). Moreover, as a careful
investigation of the Tables reveals, the jump model appears to be remarkably
stable with respect to the dataset considered, in contrast with the other
models. In particular, the values of the jump parameters vary little from one
dataset to another (the statistical significance of the fit being also somewhat
more stable). Finally it should be emphasized that, starting from a smooth
transition, the likelihood minimization procedure itself naturally leads us to a
sharp transition. It is in this sense that we view the \textbf{jump} deficit
proposal as being suggested in an especially natural way by the data.

In Fig.~\ref{fig:Pk} we show the difference between the sharp \textbf{jump}
deficit and the smooth \textbf{plaw} and \textbf{atan1} deficits, together with
the consequences for the matter power spectrum $P_\mathrm{m}(k)$.
Figure~\ref{fig:Cls} shows that, when translated into CMB anisotropies, the
\textbf{jump} deficit decreases the $C_\ell$'s at larger multipoles than is the
case for the smooth deficits \textbf{plaw} and \textbf{atan1}. Specifically,
\textbf{jump} produces a larger angular-power deficit in the region
$\ell=$20--30.

As we have noted, the case of a sharp transition exhibits an intriguing
stability across our datasets. For $C_\ell^{TT}$ only, we obtain a variation of
less than 2\% for the best-fit characteristic scale $k_\mathrm{c}^{-1}$, which
is found to range from 353 to 361 Mpc, and we find a best-fit dip in power down
to between 76 and 79\% (that is, we find $\beta$ between 0.76 and 0.79). The
$p-$value is around 3\%.

Including the polarization for the low multipoles increases the $p-$value to
around 10\%, while increasing the scale $k_\mathrm{c}^{-1}$ to about 370 Mpc and
leaving the dip $\beta$ essentially unchanged at around 0.80 or 80\%. With the
full polarization data the scale is pushed upward even more, reaching about 380
Mpc, while the dip reaches 83\% at most and the $p-$value is lowered to 7\%.
Thus it would appear from the data that we are unable to unambiguously assert
the existence of a new characteristic length scale, although our analysis
naturally points to one. Even so, as shown in Figs.~\ref{corner:full} and
\ref{corner}, this variation in the value of the scale is well within the error
bar for $k_\mathrm{c}^{-1}$.

Finally, we have computed the Bayes factor for the comparison of \textbf{jump}
and \textbf{plaw} using our full dataset (Planck $TT$, $TE$, $EE$ + lowP, H0 and
BAO-derived distances). As stated before, we use flat priors for the
cosmological parameters together with the PFI priors. For nested models, the
priors of the common parameters do not change the final Bayes factor (see for
instance Ref.~\cite{Trotta2007}). The only relevant priors are those for the new
parameters:
$$
-1 \leq \ln\left(\frac{k_\mathrm{c}^{-1}}{ 1\;\mathrm{Mpc}}\right) \leq
18,\qquad  0 \leq \beta \leq 1.
$$
Instead of reporting a single value for $B_\textsc{fa}$ for the above priors, we
adopt the framework of a robust Bayesian analysis. The prior of $\beta$ is
chosen so that all possible deficits are allowed and are equally probable.
However, there is no clear way to choose a meaningful interval for the prior of
$\ln(k_\mathrm{c})$. We circumvent this problem by calculating the Bayes factor
for all subintervals of $-1 \leq \ln\left(\displaystyle\frac{k_\mathrm{c}^{-1}}{
	1\;\mathrm{Mpc}}\right) \leq 18$ and plotting the results in Fig.~\ref{Bfactor},
thereby showing the sensitivity to the prior. The Bayes factor for the full
interval is found to be
$$ 
B_\textsc{fa} = 10^{1.06} =11.5.
$$ 
This is considered ``strong'' on the Jeffreys' scale given in Ref.~\cite[Appendix
B]{Jeffreys1998}.

In Fig.~\ref{Bfactor} there is no interval where the evidence is ``decisive'' or
``very strong'', although in a large portion of the graph the Bayes factor is
classified as ``strong''. On the other hand, the figure shows that to achieve
``strong'' evidence it is necessary to include all peaks or the last two peaks and
part of the plateau in the distribution of $k_\mathrm{c}^{-1}$ be allowed by the
prior (see Fig.~\ref{corner:full}). For priors including only one peak, the
evidence drops to ``substantial''. Thus a conservative conclusion is that we have
only ``substantial'' evidence for a particular scale (at $k_\mathrm{c}^{-1} \simeq
380\;\mathrm{Mpc}$).

\subsection{Tensor modes and scalar-tensor consistency} \label{rTS}

In our data analysis we have ignored tensor modes (effectively assuming a
negligible tensor-to-scalar ratio $r\simeq0$). It is however noteworthy that,
for both \textbf{atan} and \textbf{expc} as well as for \textbf{jump}, we have
found a general degradation of the frequentist significance when polarization
data are added (though noting that the parameters remain remarkably stable for
\textbf{jump}). The pattern of degradation depends on the datasets considered.
For example, comparing $TT$ only (Table~\ref{tab:TT_onlyMP}) with the full
Planck data (Table~\ref{tab:TTTEEE_lowPMP}) the degradation of $p-$values for
\textbf{atan} is worse than for \textbf{jump}, while comparing $TT$ + BAO (Table
\ref{tab:TT_only_BAOMP}) with $TT$ + BAO + lowP (Table~\ref{tab:TT_lowP_BAOMP})
the degradation is worse for \textbf{jump}. For $TT$ + BAO and full Planck + BAO
(Table~\ref{tab:TTTEEE_lowP_BAOMP}) the degradations are comparable. On the
other hand it should be noted that the $\Gamma$ values are persistently higher
for \textbf{jump}, indicating that the degradation is less severe. To explain
these observations, we may consider three possible scenarios.

First, we may imagine that we are in fact simply modeling a statistical fluke.
This would imply that the more data we add, the smaller the resulting
significance. However, this does not generally appear to be the case. Although
there seems to be a systematic decrease in significance of the fits when we add
polarization data, adding other datasets does not result in any particular
systematic (positive or negative) trend for the fits.

A second possibility is that our models are indeed fitting a real feature but
also partially overfitting statistical fluctuations (a combination of cosmic
variance, sample variance and data noise). In this case, adding more data may be
expected to wash out the overfit simply by reducing the variance, and this in
turn would reduce the significance to its genuine value. For each of our deficit
functions it can then happen that, because of an overfitting of the datasets
without polarization, significance is lost when polarization data are added. If
this turns out to be the true explanation for the observed degradation, then
\textbf{jump} will arguably be the preferred deficit function because of its
persistently high values of $\Gamma$ (though in terms of $p-$values it is not
clear which function would be preferred).

The third scenario, which we focus on in this section, is related to the
assumption of negligible tensor contributions. This is arguably something of a
theoretical prejudice, stemming from the $\Lambda$CDM + inflation paradigm with
a small value of $r$ (specifically $r_{0.002} < 0.099$, with 95\% CL from
temperature and polarization data), the relevant constraint being calculated
within that framework. In contrast with this paradigmatic case,
Ref~\cite{Ade:2015lrj} fitted several variants. For instance, allowing the
possibility of a running spectral index for the scalar PPS, the upper bound on
$r$ becomes $r_{0.002} < 0.15$ (95\% CL, with temperature and polarization
data). Thus changing the framework (for example allowing for a running spectral
index) permits us to relax the constraint on $r$. Including a deficit function
in the PPS can be even more drastic as it can break the inflationary
scalar-tensor consistency. In the context of our analysis it is therefore
natural to expect that a larger tensor contribution is allowed. We are then led
to consider that the addition of a tensor contribution, with a breaking of
scalar-tensor consistency, might enable us to avoid the degradation noted above.
If this turns out to be the true explanation for the observed degradation, then
\textbf{atan} will arguably be the preferred deficit function because (as
discussed below) quantum relaxation models naturally allow for a breaking of
scalar-tensor consistency.

The $TT$ angular power spectra are functionals of the scalar ($\mathcal{P}
_\textsc{s}$) and tensor ($\mathcal{P}_\textsc{t}$) power spectra:
\begin{equation}
C_{\ell}^{TT}=C_{\textsc{s},\ell}^{TT} [\mathcal{P}_\textsc{s}] +
C_{\textsc{t},\ell}^{TT}[\mathcal{P}_\textsc{t}]\ .
\end{equation}
We may parametrize $\mathcal{P}_\textsc{t}$ with an amplitude
$\mathcal{A}_\textsc{t}=r\mathcal{A}_\textsc{s}$ and write
$\mathcal{P} _\textsc{s}=\mathcal{A}_\textsc{s}f(k)$ and
$\mathcal{P}_\textsc{t}=r\mathcal{A}_\textsc{s}g(k)$. Since the
functionals are linear we have
\begin{equation}
C_\ell^{TT}=A_\textsc{s} C_{\textsc{s},\ell}^{TT} [f] + r A_\textsc{s}
C_{\textsc{t},\ell}^{TT} [g].
\label{CTT_gen}
\end{equation}
Our polarization angular power spectra are also linear functionals of
$\mathcal{P}_\textsc{s}$ and $\mathcal{P}_\textsc{t}$ and so similarly we
have
\begin{equation}
C_\ell^{XY}=A_\textsc{s} C_{\textsc{s},\ell}^{XY}
[f]+rA_\textsc{s} C_{\textsc{t},\ell}^{XY}[g],
\label{CXY_gen}
\end{equation}
where $X$ and $Y$ can denote the possible polarizations $E$ and $B$ (as
well as the temperature $T$). If $r$ is very small, the total values of
$C_\ell^{XY}$ (including $C_\ell^{TT}$) will be determined essentially
by $\mathcal{P}_\textsc{s}$ only.

Now the data seem to show anomalously low values of $C_\ell^{TT}$ for
$\ell$ roughly in the interval $[2,50]$. If we modify the function
$f(k)$ appropriately we can improve our fit to the $TT$ data in this
region. However, it is not so straightforward if we consider a dataset
that includes polarization. For example, if we take the datasets $TT$
and $TE+EE$, the component $TE+EE$ does not have the same anomalously
low power at low $\ell$. Thus if we try modifying $f(k)$ so as to
improve the fit to the $TT$ data, at some point we will worsen the fit to
the $TE+EE$ data. In other words, with a total likelihood
\begin{equation}
\ln L_\mathrm{total} = \ln L_{TT}[A_\textsc{s},f] + \ln
L_{TE+EE}[A_\textsc{s},f],
\end{equation}
lowering the power in $f$ will increase the first term but decrease the
second, so that the best fit will be somewhere in the middle ground.

If instead the tensor contribution is not negligible on all scales of
interest, it may be possible to increase $\ln
L_{TT}[A_\textsc{s},r,f,g]$ without decreasing $\ln
L_{TE+EE}[A_\textsc{s},r,f,g]$ -- provided we are allowed to vary the
functions $f$ and $g$ independently. For then we might be able to choose
$f$ such that we have lower power in $C_\ell^{TT}$ on large scales while
at the same time choosing $g$ such that it compensates for the lower
power in $C_\ell^{TE}$ and $C_\ell^{EE}$ at large scales (resulting from
lower values of $C_{\textsc{s},\ell}^{TE}[f]$ and
$C_{\textsc{s},\ell}^{EE}[f]$) and without spoiling the rest of the fit.
This would require the relative tensor contributions to $TE$ and $EE$ to
be larger than the relative tensor contribution to $TT$, which can occur
in appropriate conditions.

However, to vary $f$ and $g$ independently amounts to a violation of
scalar-tensor consistency. In such conditions the definition
\begin{equation}
r\equiv4\frac{\mathcal{P}_\textsc{t}}{\mathcal{P}_\textsc{s}}\label{r}
\end{equation}
of $r$ is no longer a fixed number but will generally depend on $k$. In
practice, however, $r$ is taken to be the ratio (\ref{r}) at the pivot
scale $k=k_{\ast}$ (with $\mathcal{A}_\textsc{s}$ defined as the value
of $\mathcal{P}_\textsc{s}$ at $k=k_{\ast}$ so that $f=1$ at that
point), in which case the general relations (\ref{CTT_gen}) and
(\ref{CXY_gen}) are still valid. Note that while this scenario requires
a large contribution from tensor modes at large scales, $r$ itself could
still be small since it is defined at the relatively small pivot scale.

This reasoning suggests that, if the low-power anomaly in $C_\ell^{TT}$
is real, then having significant tensor contributions at large scales
(with a violation of scalar-tensor consistency) might allow us to avoid
a degradation of the fit when polarization data are included.

An intriguing feature of quantum relaxation models is that they
naturally imply a large-scale violation of scalar-tensor consistency
without spoiling the overall inflationary scenario
\cite{Valentini:2008dq}. This is because, when the initial conditions
are no longer constrained by the Born rule, there is no reason why
different degrees of freedom should have the same initial nonequilibrium
distribution and hence there is no reason why they should have the same
large-scale deficit function $\xi(k)$.\footnote{See
Ref.~\cite{Colin:2015tla} for examples where different initial
nonequilibrium distributions all give rise to approximately
\textbf{atan} spectra but with different parameter values.} In general
we will have two distinct functions $\xi_\textsc{s}(k)$ and
$\xi_\textsc{t}(k)$ for scalar and tensor modes respectively, with two
different and unequal sets of parameters $\alpha_\textsc{s}$,
$\beta_\textsc{s}$, $(k_{\mathrm{c}})_\textsc{s}$ and
$\alpha_\textsc{t}$, $\beta_\textsc{t}$, $(k_{\mathrm{c}})_\textsc{t}$
(with $\lambda=1$ throughout). A more complete data analysis would then
require a fit to this six-parameter model, where for completeness $r$
itself could also be subject to a fresh fit. Such studies are left for
future work.

\section{Implications for quantum relaxation models} \label{sec:QR}

In this section we consider the implications of our data analysis for
quantum relaxation models
\cite{Colin:2014pna,Colin:2015tla,Colin:2013rwa,Valentini:2008dq}.

\subsection{Best-fit results for the nonequilibrium deficit}

As noted in Sec.~\ref{sec:deficit}, quantum relaxation during a
radiation-dominated preinflationary era (combined with a simplifying assumption
about the transition to inflation) predicts an approximate deficit function
$\xi_{\mathrm{neq}}(k)$ of the form \eqref{xineq0} with $\lambda=1$
\cite{Colin:2014pna,Colin:2015tla}. That is, the relaxation scenario predicts
the deficit function that we have here called \textbf{atan3}, with three
undetermined parameters $\alpha$, $\beta$ and $k_{\mathrm{c}}$. In the present
analysis we have fitted the data to \textbf{atan3}, and also to the reduced
functions \textbf{atan2} (with $\alpha=1$) and \textbf{atan1} (with $\alpha=1$
and $\beta=0$). The results show that conclusions about the fits depend on the datasets considered, in particular on whether we consider only datasets with no polarization (Tables
\ref{tab:TT_onlyMP}--\ref{tab:TT_only_H0_BAOMP}) or whether we instead consider only datasets that include polarization (Tables
\ref{tab:TT_lowPMP}--\ref{tab:TTTEEE_lowP_H0_BAOMP}).

As a general point of principle, it could happen that a useful pattern emerges
only for datasets that include polarization. More complete datasets can be
required to observe an effect, where more data generally implies smaller error
bars (by diminishing the data variance). In this spirit it may be useful to
consider Tables \ref{tab:TT_onlyMP}--\ref{tab:TT_only_H0_BAOMP} and Tables
\ref{tab:TT_lowPMP}--\ref{tab:TTTEEE_lowP_H0_BAOMP} as two distinct sets of
data.

\subsubsection{Best fit without polarization}

If we restrict ourselves to datasets without polarization, for a reliable best
fit \textbf{atan1} does not suffice and we require \textbf{atan2} or
\textbf{atan3}.

To see this, observe that in Tables
\ref{tab:TT_onlyMP}--\ref{tab:TT_only_H0_BAOMP} the parameter
$k_{\mathrm{c}}$ changes considerably from \textbf{atan1} to
\textbf{atan2} (where the latter fit yields quite a large
$\beta\simeq0.5$), showing that \textbf{atan1} is not a reliable fit.
Thus, while \textbf{atan1} shows an apparently impressive significance
of up to $2.7\sigma$ (Table~\ref{tab:TT_only_H0MP}) for these data
sets, the instability of the fit indicates that this result should be
discounted. Whereas, again for Tables
\ref{tab:TT_onlyMP}--\ref{tab:TT_only_H0_BAOMP}, the parameters
$k_{\mathrm{c}}$ and $\beta$ are more or less stable from \textbf{atan2}
to \textbf{atan3} (although less stable for Table
\ref{tab:TT_only_H0_BAOMP}), suggesting that \textbf{atan2} is a
reliable fit -- with a significance of up to $2.4\sigma$ (Table
\ref{tab:TT_only_H0MP}). The latter result is suggestive, but as we
shall discuss the significance diminishes when polarization data are
included.

\subsubsection{Best fit including polarization}

If instead we ignore datasets without polarization, we find that the only
relevant parameter is the (uncertain) scale $k_{\mathrm{c}}$, so that the
best-fit function is in effect just \textbf{atan1}. However, we cannot really
conclude that $\beta=0$ because the data cannot provide such a constraint. In
other words, the likelihood (for \textbf{atan2} and \textbf{atan3}) is almost
constant around $\beta = 0$ (see Tables
\ref{tab:TT_lowPMP}--\ref{tab:TTTEEE_lowP_H0_BAOMP}). This is a direct
consequence of $\beta$ modifying the spectrum only at values of $k^{-1}$ much
larger than $k_\mathrm{c}^{-1}$ (as discussed in Sec.~\ref{app:atan}): since
$k_\mathrm{c}^{-1}$ is already large, $\beta$ modifies the spectrum at
unobservable scales. To obtain an upper bound on $\beta$, we could vary the
initial $\beta$ values in the fitting process making them closer and closer to
one (or we could perform a full MCMC posterior analysis). For the purpose of
model comparison, however, it suffices to obtain the fits with $\beta \approx
0$.

\subsubsection{Significance of atan1}

If we consider all the datasets, we find, as a rough general trend, that the
more data we add the larger the best-fit lengthscale $k_{\mathrm{c}}^{-1}$ and
the smaller the significance of the fit. This suggests that the effect is
probably a statistical fluctuation. In principle, however, it could be that the
effect only occurs at super-Hubble scales  and that the \textbf{atan1} model is
trying to fit a real physical feature there. With this in mind, if we allow
ourselves to disregard the datasets without polarization (and if we fix
$\lambda=1$ as in the quantum relaxation model), then it is reasonable to
consider only \textbf{atan1} since at these scales neither $\alpha$ nor $\beta$
have a measurable impact on the spectrum. Thus \textbf{atan1} may be regarded as
a smooth alternative to the sharp function \textbf{jump}. For datasets
including polarization, \textbf{atan1} is found to have a significance of up to
only $2\sigma$ (Table \ref{tab:TT_lowP_H0_BAOMP}).

\subsubsection{Degradation and stability of fits. Comparison with \textbf{jump}}

From the point of view of significance \textbf{atan1} performs about as
well as \textbf{jump} (overall for datasets including polarization
both at low- and high-$\ell$), having in some cases a slightly better
significance for the same dataset. However, the significance of
\textbf{atan1} is found to diminish systematically as more polarization
data are included (see Table XIII), that is, including polarization at
low-$\ell$ decreases the significance and when both low- and
high-$\ell$ are included the significance decreases even further. By
contrast, while the significance of \textbf{jump} decreases when
low-$\ell$ polarization is added it increases again for the full
polarization data (see Table XIV). Both deficit functions have a
significance varying from $\approx1.5\sigma$ to $\approx2.5\sigma$.

Regarding the general stability of the fits, as discussed in Ref.~\cite{KV19}
there are different regions of the \textbf{atan} parameter space that produce
approximately the same curve. Such degeneracy can result in a large variation
and apparent instability of the best-fit parameters for \textbf{atan}. In
contrast, for \textbf{jump} there can be no such degeneracy. As we have seen the
parameters for \textbf{jump} are found to be stable, and this of course implies
that the function itself is stable. This fact, together with the larger values
of $\Gamma$ for \textbf{jump}, motivated our follow-up Bayesian analysis carried
out above. The case for running a similar Bayesian analysis for \textbf{atan} is
not as strong: the stability of the function remains to be clarified as does the
structure of the parameter space, and in any case the values of $\Gamma$ are
lower. Thus we leave such further analysis for future work.

\subsection{Quantum relaxation and future work}

We now comment on the implications of our results for future work on quantum
relaxation models.

\subsubsection{Mechanism for negligible super-Hubble power}

As far as $p-$values are concerned, \textbf{atan1} performs more or less
as well as \textbf{jump} (for datasets including polarization). This
motivates us to ask if there might be a theoretical model that predicts
\textbf{atan1} and in particular a near-vanishing $\beta$. Because
$\beta=\lim_{k\rightarrow0} \xi_{\mathrm{neq}}(k)$, this means that we
require a model in which the primordial power spectrum itself becomes
negligible in the far super-Hubble limit.

In a quantum relaxation scenario with a radiation-dominated
preinflationary phase, the limit $k\rightarrow0$ yields the maximum
suppression or ``freezing'' of quantum relaxation. As shown in Sec.~V
of Ref. \cite{Colin:2013rwa}, in the far super-Hubble regime the de
Broglie-Bohm time evolution of a field mode on an interval
$(t_\mathrm{i},t_\mathrm{f})$ with $t_\mathrm{f}\gg t_\mathrm{i}$ is
equivalent to the time evolution of a standard harmonic oscillator on a
time interval $(t_\mathrm{i} ,3t_\mathrm{i})$. Thus for $k\rightarrow0$
all modes effectively evolve over the same small time $2t_\mathrm{i}$
and we expect very limited relaxation. At small $k$ the resulting
deficit function $\xi_{\mathrm{neq}}(k)$ will then be essentially equal
to the deficit function $\xi_{\mathrm{ic}}(k)$ associated with the
initial conditions. This means that, for long-wavelength modes, the
freezing of relaxation preserves the initial conditions almost intact. A
negligible value of $\beta$ then implies a negligible statistical
variance (or power) in the initial conditions themselves (for modes in
the far super-Hubble regime).

This motivates us to consider a quantum relaxation scenario in which
there is negligible super-Hubble noise in the initial state
(corresponding to very small $\beta$). It is a matter for future
theoretical work to develop the details of such a scenario. A simple
suggestion is to assume that there is negligible power in the initial
conditions for all modes. This is an attractive hypothesis, as
physically it means that essentially all of the quantum noise we observe
at later times was generated dynamically\footnote{In Ref. \cite[Sec.
X]{Valentini:2008dq} it was argued that, in a theory of dynamical
quantum relaxation, it is natural to have an initial subquantum width
(so that $\xi_{\mathrm{ic}}(k)<1$). Following the same logic to its
conclusion, it is arguably natural to take $\xi_{\mathrm{ic}}(k)$ to be
as small as possible and to set $\xi _{\mathrm{ic}}(k)\simeq0$.}.

While the hypothesis of negligible initial power provides a good
physical reason to prefer \textbf{atan1}, so far the significance
remains small. A full MCMC analysis might help to evaluate whether or
not it is worth pursuing such models. This would depend on the resulting
upper bound on $\beta$.

If this hypothesis is considered further, it would be natural to apply
the same reasoning to tensor modes as well, in which case we would
expect the distinct functions $\xi_\textsc{s}(k)$ and
$\xi_\textsc{t}(k)$ (discussed in Sec.~\ref{rTS}) to each take the
form of \textbf{atan1} but with different scales
$(k_{\mathrm{c}})_\textsc{s}$ and $(k_{\mathrm{c}})_\textsc{t}$ (again
generally breaking scalar-tensor consistency).

\subsubsection{Other signatures of quantum relaxation}

The data seem to show that if there is a low-power anomaly it must exist
at large scales that we cannot accurately measure. Because cosmic
variance is so large in the relevant region, we are unable to
meaningfully test the predictions of quantum relaxation for the power
deficit alone. To improve the chances of constraining such models we
need to include more detailed predictions$-$such as primordial
oscillations and statistical anisotropy, which are additional generic
features of quantum relaxation. Extensive numerical simulations show
significant oscillations around the \textbf{atan} function
\cite{Colin:2014pna,Colin:2015tla}, which have however so far eluded a
simple and general parametrization. Statistical anisotropy arises from
initial nonequilibrium distributions that depend on the direction
$\bm{\hat{k}}$ of the wave vector, resulting in parameters $\alpha$,
$\beta$, $k_{\mathrm{c}}$ that depend on $\bm{\hat{k}}$, where the
effect of $\alpha(\bm{\hat{k}})$ could arguably persist at large $\ell$
and hence have a more visible impact on the data
\cite{Valentini:2015sna}.

\subsubsection{Quantum relaxation across the transition}

Finally, it should be emphasised that the \textbf{atan} prediction was
obtained on the simplifying assumption that the transition does not affect
the nonequilibrium distribution left at the end of a radiation-dominated
preinflationary era \cite{Colin:2014pna}. Modeling the transition and
simulating the time evolution of nonequilibrium across it remains to be
done. How this might change the overall result is currently unknown. The
evidence discussed above in favor of a sudden \textbf{jump} deficit
raises the question of whether or not a realistic quantum relaxation
model could yield such a result (or, indeed, if such a result could
arise from other kinds of models not involving quantum relaxation).

\section{Conclusions} \label{sec:conc}

A smooth deficit function can be superimposed on the primordial power spectrum
to mimic the large-scale deficit which has apparently been observed in some
cosmological data. We have analyzed a broad range of data using different
parametrized versions of the deficit function, in such a way as to be able to
compare with a previous analysis by the Planck team. We confirm that, for the
deficit functions we consider, the fit is only marginally better than for the
standard power law, where the improvement occurs at wavelengths comparable to
the Hubble scale. It would appear that, taken by itself, the power deficit is
not very statistically significant and therefore not necessarily physical. This
result is consistent with previous investigations. We do, however, find some
suggestive hints for future work.

We have consistently found hints of statistically-significant fits, only to find
that the significance degrades when polarization data are added. We have argued
that such degradation might be avoided in models that break scalar-tensor
consistency and which have non-negligible tensor contributions at large scales.
Quantum relaxation models in fact naturally break scalar-tensor consistency,
yielding independent deficit functions for scalar and tensor degrees of freedom.
Fitting such extended models to the data may be considered in future work.
Another possibility, however, is that our fits without polarization are
partially overfitting those datasets, so that when polarization data are added
this part of the modeling loses its significance.

The behavior of the restricted (one-parameter) quantum relaxation
deficit function \textbf{atan1} across all datasets arguably suggests
that it is merely modeling a statistical fluctuation, since adding more
data tends to increase the lengthscale $k_{\mathrm{c}}^{-1}$ and
decrease the significance. Possibly, however, the fit is trying to
capture a real feature at super-Hubble scales. To test this, we might
consider disregarding the datasets without polarization, in which case
our results do suggest that \textbf{atan1} may be worth considering
further, in particular because the data seem to be relatively
insensitive to the values of the additional parameters $\alpha$ and
$\beta$. Physically, the function \textbf{atan1} has vanishing power in
the far super-Hubble limit, and we have argued that this would be a
natural feature in quantum relaxation models with negligible power in
the initial conditions.

Future theoretical work on the power deficit should, however, also take
note of the following elementary point. Because of the low statistical
significance of the deficit, an effective test of theoretical models
will require that we include further predictions such as primordial
oscillations or statistical anisotropy, especially if these are able to
affect the data at larger values of $\ell$.

Our study of smooth deficit functions has, however, already led us to an
unexpected and statistically significant result of another kind. By
allowing the fit to run on the characteristic deficit speed, we have
found that the additional power index$-$the parameter $\lambda$ in
\eqref{xineq}$-$is naturally driven to very large values, implying an
almost discontinuous or steplike deficit function. After having scanned
much of the parameter space, we studied the specific case which the data
seemed to be pointing to: a deficit function \textbf{jump} with only two
parameters, specifically a break point $k_\mathrm{c}$ indicating the
scale above which the usual fiducial power spectrum is valid and a
relative amplitude difference $\beta$. Running our analysis with this
two-parameter step function \eqref{jump}, we obtained a fit with better
agreement with the full range of datasets (better in the sense of more
stable parameters and higher values of $\Gamma$), exhibiting a new
length scale $D_\mathrm{c}$ around $2\pi\times 350$~Mpc~$\approx
2200$~Mpc today and with a power deficit of about 20\%. This stability
indicates that the model is not merely overfitting a particular dataset,
and that the feature it fits is real. The resulting modification of the
primordial power spectrum and its impact on the matter power spectrum
are shown in Fig.~\ref{fig:Pk}. In our Bayesian follow-up analysis we
obtained ``strong'' evidence when allowing the scale to vary in a wide
interval and ``substantial'' evidence for our peak at around 350~Mpc.

Taking $D_\mathrm{c}$ at face value today, and running it backwards by
some appropriate number $N$ of e-folds to an inflationary phase during
which it may have been generated, we find a corresponding primordial
scale around $\ell_\mathrm{c} \sim 3\times 10^8
\mathrm{e}^{120-N}\ell_\mathrm{p}$, with $\ell_\mathrm{p}$ the Planck
length. For the commonly quoted value $N = 120$ (including the later
radiation- and matter-dominated epochs), in terms of energy this scale
corresponds to $\approx 2.5\times 10^{20}$~eV. If $N$ is allowed to
range up to $\sim 140$, the scale approaches $\ell_\mathrm{p}$ or an
energy scale $\sim 10^{19}$~GeV.

\begin{figure}
\centering \includegraphics[scale=1.0]{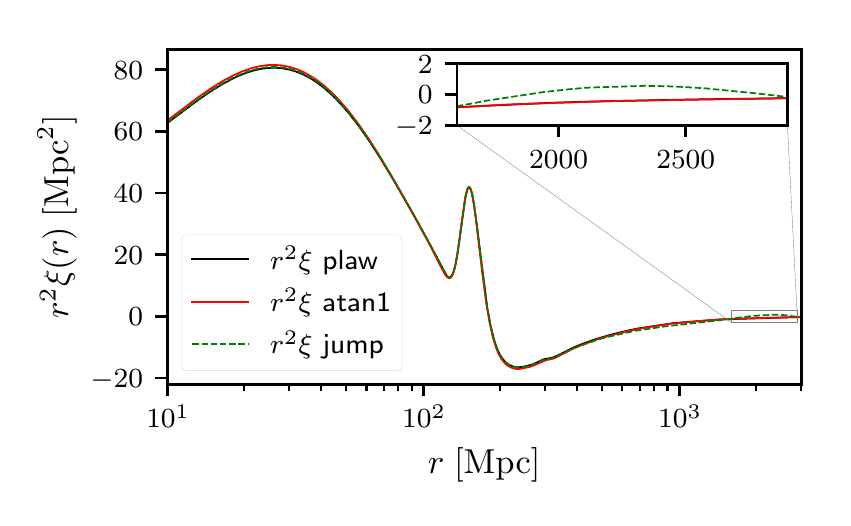}
\caption{The predicted linear	(3D) two-point functions for perturbations
in the total cosmological matter density (conventionally denoted $\xi$ )
when the power-law primordial spectrum is modified by our best-fit
deficit functions \textbf{atan1} and \textbf{jump}. These fits are for
the full Planck dataset.\label{fig:2ptfunct}}
\end{figure}

Forthcoming experiments may yield further insights into the magnitude,
statistical significance, and physical relevance of this potentially
new scale. We may for example consider how our best-fit primordial
deficit function \textbf{jump} would affect the two-point correlation
function (in three-dimensional space) for perturbations in the total
cosmological matter density, as traced by the distribution of galaxies.
This is shown in Fig.~\ref{fig:2ptfunct}. The \textbf{jump} function
creates a very small bump at $r \approx 2 \;\mathrm{Gpc}$, which is
compatible with our predicted scale $2\pi k_\mathrm{c}^{-1} \approx
2200\;\mathrm{Mpc}$. A bump at such a scale might be observable in
upcoming surveys. This is demonstrated by (say) the BOSS
results~\cite{Laurent:2016eqo,Ntelis:2017nrj}, which come close to
measuring this scale using a $3 (\mathrm{Gpc/h})^3$ volume for the
galaxy sample and a $14 (\mathrm{Gpc/h})^3$ value for the quasar
sample. While it is clear that BOSS is not able to resolve such large
scales, Euclid (for example) might possibly be able to do so at least
partially. Euclid~\cite{2011arXiv1110.3193L} will increase this volume
to $\sim 50$~Gpc$^3$ which could observe (at least partially) the
relevant scales. Of course one must include other effects, such as
redshift-space distortions, in order to be able to compare with actual
data, and such effects could make this bump difficult to measure even
if it exists. Even so, a possible empirical confirmation of this
potentially new scale seems within reach.

\acknowledgments

This work is based on observations obtained with Planck
(\url{http://www.esa.int/Planck}) and the likelihood code \textsc{plc}
from the Planck Legacy Archive. Planck is an ESA science mission with
instruments and contributions directly funded by ESA Member States,
NASA, and Canada.

P.~P. and S.~D.~P.~V. would like to thank the Labex Institut Lagrange de
Paris (reference ANR-10-LABX-63), part of the Idex SUPER, within which
this work was partly done. S.~D.~P.~V. would like to acknowledge
financial aid from the CNPq PCI/MCTI/CBPF program, PNPD/CAPES (Programa
Nacional de Pós-Doutorado/Capes, reference 88887.311171/2018-00)  and
financial support from a BELSPO non-EU postdoctoral fellowship. P.~P. is
hosted at Churchill College, Cambridge, where he is partially supported
by a fellowship funded by the Higher Education, Research and Innovation
Dpt of the French Embassy to the United-Kingdom. A.~V. is grateful to
Murray Daw and Mark Leising at Clemson University for their support
during this project.

This research also made use of the MeSU supercomputer of the Universit\'{e}
Pierre \& Marie Curie and the Horizon Cluster funded by the Institut
d'Astrophysique de Paris. We thank St\'ephane Rouberol for running this cluster
smoothly for us.

Finally, we would also like to thank R.~Trotta for a careful reading of
the manuscript and for many suggestions for improvements, as well as
B.~Sherwin and W.~Handley for enlightening discussions.

\appendix

\section{NUMCOSMO}
\label{app:numcosmo}

In this appendix we briefly describe the numerical tools used in this
work. These tools are part of the Numerical Cosmology library --
N\textsc{um}C\textsc{osmo}~\cite{Vitenti2012c}. All codes described here
are located in the project's
repository~\url{https://github.com/NumCosmo/NumCosmo} and the library's
documentation is in \url{https://numcosmo.github.io/}. The library
contains an independent implementation of several tools used in
numerical cosmology, providing a complete toolkit to compute and analyze
different cosmological observables. The observables were computed with
the homogeneous and isotropic cosmological models objects
(N\textsc{c}HIC\textsc{osmo}*) and with the Boltzmann code using
C\textsc{lass}~\cite{Blas2011,Lesgourgues2011a,Lesgourgues2011b} as
back-end. Different precision files were used, all based on
\textsf{chi2pl0.01.pre} (which was calibrated by the C\textsc{lass}
developers to provide a $10^{-3}$ error in $C_\ell$ at version 2.5.0).
This precision file was modified to increase the number of points per
decade to evaluate the PPS and to decrease the distance between
interpolation points in $\ell$ (no interpolation for low-$\ell$ and
increasing interpolation for high-$\ell$). This was necessary to capture
the features added by the different deficit functions. We used three
different precisions:
\begin{itemize}
	\item[LP:] The default class precision;
	\item[MP:] The original file \textsf{chi2pl0.01.pre} with the addition -- \textsf{k\_per\_decade\_primordial} = 50.
	\item[HP:] Same as MP but using -- \\
	\textsf{l\_logstep} = 1.02,\\
	\textsf{q\_linstep} = 0.2.
\end{itemize}
After running the best-fit finder using these precisions, we found no relevant
differences between MP and HP.

For the best-fit finders we used NumCosmo's \textsf{NcmFit} object and the NLOpt
(see Ref.~\cite{Johnson2014}) library as the minimization library back-end. We
tested different numerical optimization algorithms. The most efficient and
stable algorithms found were the Nelder-Mead~\cite{Nelder1965, Box1965,
	Richardson1973} and Subplex~\cite{Rowan1990}. The main advantage of these
algorithms is that they do not require the objective function derivatives and
have a better handling for discontinuous functions. The likelihood used is, in
principle, a smooth function of the parameters $\theta$, although the
computation of $C^{XY}_\ell$ introduces numerical errors in the evaluation of
$L(D|\theta)$, which in turn can create artificial discontinuities at the error
level. Accordingly, any optimization algorithm that depends on the smoothness of
the likelihood (for example, some algorithms create a cubic approximation to the
objective function or calculate an estimate of the derivative through finite
differences) will find spurious maxima resulting from these small
discontinuities (mainly on regions/direction where the likelihood is almost
constant). In our fitting process we found that even the more appropriate
algorithms terminate prematurely due to the artificial maxima. For this reason,
we rerun the fitting process iteratively until the last two minima coincide
within a 0.1\% margin. This process was automated in the
\textsf{NcmFit.run\_restart} method.

We note a small difference between the parameter best-fit values using
different precisions for the CMB anisotropies and/or accuracy for the
minimization algorithm. The code used by Planck,
C\textsc{osmo}MC~\cite{Lewis2000, Lewis2013}, uses the
BOBYQA~\cite{Powell2009} (Bound Optimization BY Quadratic Approximation)
as the best-fit finder or even the point with smaller $-2\ln(L)$ value
found during the MCMC exploration. We also tested these two methods.
Both provide suboptimal minima with small differences in the best-fit
parameters. The higher precision for the computation of the
$C^{XX}_\ell$ and a more precise minimization algorithm resulted in
best-fit parameters close to the ones previously obtained by the Planck
team (with differences of the order of the precision for the
$C^{XX}_\ell$, $\approx 10^{-3}$) but with larger differences in the
values of the likelihood itself, that is, in the values of $\Gamma_i$
(with $\Delta\Gamma \approx 0.5$).

The MCMC algorithm used is identical to that used in Appendix D of
Ref.~\cite{Doux2018}. Note however that, in our analysis, we did not use
the profile likelihood for the PFI parameters: instead, we used a
complete sample including all parameters. In Appendix E of
Ref.~\cite{Doux2018}, one can find the description of all the
diagnostics used to assert the convergence of the MCMC sampler.

Finally, we used the harmonic mean (see for context~\cite{Weinberg2012})
to estimate the evidence integral. In our implementation, our posterior
sample was obtained from the MCMC to compute the integral
\begin{equation*}
\left\langle g(\theta) \right\rangle = \int\dd\theta
g(\theta)P(\theta\vert\mathcal{M},D) \approx \frac{1}{N} \sum_i
g(\theta_i),
\end{equation*}
where $\theta_i$ are the $N$ points in the posterior MCMC sample. Using
Bayes theorem we have
\begin{equation*}
\int\dd\theta
g(\theta)\frac{P(\theta\vert\mathcal{M})
L(D\vert\theta,\mathcal{M})}{P(D\vert\mathcal{M})} 
\approx \frac{1}{N} \sum_i g(\theta_i),
\end{equation*}
where $L(D\vert\theta,\mathcal{M})$ is the likelihood and
$P(\theta\vert\mathcal{M})$ the priors. We usually do not know the
properly normalized likelihood $L(D\vert\theta,\mathcal{M})$, thus we
define $L^\prime(D\vert\theta,\mathcal{M}) \equiv N_L
L(D\vert\theta,\mathcal{M}) $, as the unnormalized likelihood (which is
used in most cases for MCMC), and $N_L$ the normalization factor (which
does not depend on any parameter), where the value of
$P(\theta\vert\mathcal{M})L^\prime(D\vert\theta,\mathcal{M})$ was the
one used by the MCMC sampler. For this reason we can choose
$$
g(\theta) = \frac{F(\theta)}{P(\theta\vert\mathcal{M})
L^\prime(D\vert\theta,\mathcal{M})},
$$
to obtain
\begin{equation*}
\frac{1}{N_LP(D\vert\mathcal{M})}\int\dd\theta F(\theta) \approx
\frac{1}{N} \sum_i \frac{F(\theta_i)}{P(\theta_i\vert\mathcal{M})
L^\prime(D\vert\theta_i,\mathcal{M})}.
\end{equation*}
We then choose $F(\theta)$ to be a multivariate normal distribution with
the mean and covariance equal to the estimates from the MCMC sample
$\theta_i$. However, the support of $F(\theta)$ must be the same as that
of $P(\theta\vert\mathcal{M},D)$ and therefore we normalize $F(\theta)$
in the same domain as $P(\theta\vert\mathcal{M},D)$ obtaining a
truncated multivariate normal distribution. After the normalization we
finally get our estimator for the reciprocal of $P(D|\mathcal{M})$
(modulo the irrelevant constant factor $N_L$):
\begin{equation*}
\hat{\mathcal{Z}} \equiv \frac{1}{N_LP(D\vert\mathcal{M})} \approx
\frac{1}{N} \sum_i \frac{F(\theta_i)}{P(\theta_i\vert\mathcal{M})
L^\prime(D\vert\theta_i,\mathcal{M})}.
\end{equation*}
We computed the error in $\hat{\mathcal{Z}}$ using two methods. The
first consists in splitting the sample $\theta_i$ into $M$ subsamples,
computing $\hat{\mathcal{Z}}_i$ for each one, and from the results
computing estimators for the mean and variance. The second method
consists in bootstrapping $\theta_i$ (re-sampling with replacement),
computing $\hat{\mathcal{Z}}_i$ for each bootstrap realization, and
applying the usual mean and variance estimators. Both methods give
numerically equivalent results.

\section{MODEL TABLES}\label{App:models}

In this Appendix, we discuss the information already present in
Tables~\ref{tab:TT_onlyMP} to \ref{tab:TTTEEE_lowP_H0_BAOMP} but
presented so as to allow an easy model-by-model comparison with respect
to various datasets. Focusing on the favored models {\bf atan1} and
{\bf jump}, we briefly discuss the performance of each model with
respect to all datasets used.

\begin{table*}
\begin{ruledtabular}
\begin{tabular}{ l | c | c | c | c | c | c | c | c | c | c | c | c | r }
data-set & $k_\mathrm{c}^{-1} [\mathrm{Gpc}]$ & $\lambda$ & $\beta$ & $\alpha$ & $h$ & $\ns$ & $\ln(10^{10}\As)$ & $\tau$ & $\Gamma$ & $\gamma$ \\
\hline 	Planck $TT$  &  $ 3.6 $  &  $ 1 $  &  $ 0 $  &  $ 1 $  &  $ 0.7 $  &  $ 0.97 $  &  $ 3.3 $  &  $ 0.18 $  &  $ 5.5 $  &  $  1.9 $\% ($ 2.4 \sigma$) $[1]$\\
\hline 	Planck $TT$+H0  &  $ 3 $  &  $ 1 $  &  $ 0 $  &  $ 1 $  &  $ 0.71 $  &  $ 0.98 $  &  $ 3.4 $  &  $ 0.21 $  &  $ 7.1 $  &  $  0.79 $\% ($ 2.7 \sigma$) $[1]$\\
\hline 	Planck $TT$+BAO  &  $ 3.5 $  &  $ 1 $  &  $ 0 $  &  $ 1 $  &  $ 0.69 $  &  $ 0.97 $  &  $ 3.3 $  &  $ 0.18 $  &  $ 4.6 $  &  $  3.2 $\% ($ 2.1 \sigma$) $[1]$\\
\hline 	Planck $TT$+H0+BAO  &  $ 3.9 $  &  $ 1 $  &  $ 0 $  &  $ 1 $  &  $ 0.69 $  &  $ 0.97 $  &  $ 3.3 $  &  $ 0.17 $  &  $ 5.6 $  &  $  1.8 $\% ($ 2.4 \sigma$) $[1]$\\
\hline 	Planck $TT$+lowP  &  $ 5.3 $  &  $ 1 $  &  $ 0 $  &  $ 1 $  &  $ 0.68 $  &  $ 0.96 $  &  $ 3.2 $  &  $ 0.11 $  &  $ 3.5 $  &  $  6.2 $\% ($ 1.9 \sigma$) $[1]$\\
\hline 	Planck $TT$+lowP+H0  &  $ 5.6 $  &  $ 1 $  &  $ 0 $  &  $ 1 $  &  $ 0.69 $  &  $ 0.97 $  &  $ 3.2 $  &  $ 0.11 $  &  $ 3.8 $  &  $  5.1 $\% ($ 1.9 \sigma$) $[1]$\\
\hline 	Planck $TT$+lowP+BAO  &  $ 5.3 $  &  $ 1 $  &  $ 0 $  &  $ 1 $  &  $ 0.68 $  &  $ 0.96 $  &  $ 3.2 $  &  $ 0.11 $  &  $ 3.5 $  &  $  6 $\% ($ 1.9 \sigma$) $[1]$\\
\hline 	Planck $TT$+lowP+H0+BAO  &  $ 5.7 $  &  $ 1 $  &  $ 0 $  &  $ 1 $  &  $ 0.69 $  &  $ 0.97 $  &  $ 3.1 $  &  $ 0.1 $  &  $ 3.9 $  &  $  4.9 $\% ($ 2 \sigma$) $[1]$\\
\hline 	Planck $TT$, $TE$, $EE$+lowP  &  $ 6.5 $  &  $ 1 $  &  $ 0 $  &  $ 1 $  &  $ 0.67 $  &  $ 0.96 $  &  $ 3.1 $  &  $ 0.094 $  &  $ 3.4 $  &  $  6.7 $\% ($ 1.8 \sigma$) $[1]$\\
\hline 	Planck $TT$, $TE$, $EE$+lowP+H0  &  $ 7.3 $  &  $ 1 $  &  $ 0 $  &  $ 1 $  &  $ 0.68 $  &  $ 0.96 $  &  $ 3.1 $  &  $ 0.096 $  &  $ 3.2 $  &  $  7.4 $\% ($ 1.8 \sigma$) $[1]$\\
\hline 	Planck $TT$, $TE$, $EE$+lowP+BAO  &  $ 7 $  &  $ 1 $  &  $ 0 $  &  $ 1 $  &  $ 0.68 $  &  $ 0.96 $  &  $ 3.1 $  &  $ 0.095 $  &  $ 3.2 $  &  $  7.4 $\% ($ 1.8 \sigma$) $[1]$\\
\hline 	Planck $TT$, $TE$, $EE$+lowP+H0+BAO  &  $ 6.6 $  &  $ 1 $  &  $ 0 $  &  $ 1 $  &  $ 0.68 $  &  $ 0.96 $  &  $ 3.1 $  &  $ 0.099 $  &  $ 3.2 $  &  $  7.1 $\% ($ 1.8 \sigma$) $[1]$\\
\end{tabular}
\end{ruledtabular}
\caption{\label{tab:atan1} Excerpt from Tables~\ref{tab:TT_onlyMP} to \ref{tab:TTTEEE_lowP_H0_BAOMP} selecting the results for \textbf{atan1}, see Table~\ref{tab:TT_onlyMP} for more details.}
\end{table*}

\subsection{Inverse tangent model: \textbf{atan1}}
\label{app:atan}

The \textbf{atan} model given by Eq.~\eqref{xineq} provides a very
smooth deficit correction to the fiducial PPS. In fact, the parameter
$k_\mathrm{c}$ is not really a characteristic scale in this case, as can
be seen by the following example: setting $\beta = 0$, $\alpha = 1$ and
$\lambda = 1$, one obtains the following values for $X_\xi \equiv
k/k_\mathrm{c}$
\begin{align}
\label{X1} X_{1\%}=0.01, \\ 
\label{X5} X_{5\%}=0.07, \\ 
\label{X20} X_{20\%}=0.33, \\ 
\label{X80} X_{80\%}=4.3, \\ 
\label{X95} X_{95\%}=19, \\
\label{X99} X_{99\%}=99,
\end{align}
where the index indicates the amplitude reduction compared
to the fiducial PPS.

In words, Eqs.~\eqref{X1} to \eqref{X99} show that $\xi_\mathrm{neq}$
starts decreasing the amplitude of the PPS already for, say,
$X_{99\%}=99$, which implies $k_{99\%}^{-1} = k_\mathrm{c}^{-1}/99$.
Therefore, for a fit with $k_\mathrm{c}^{-1} \approx 3.5 \;\mathrm{Gpc}$
say, the modification starts around $\approx 35\; \mathrm{Mpc}$, the
power drop reaching 20\% for $k_{80\%}^{-1} = k_\mathrm{c}^{-1}/4.3
\approx 914\;\mathrm{Mpc}$. Moreover, the total deficit window can be
very wide: in the case at hand, one has $X_{99\%}/X_{1\%} \approx 10^4$,
so that $k$ varies over four orders of magnitude within the deficit
window. A more conservative window, from 20\% to 80\% say, leads to
$X_{80\%}/X_{20\%} \approx 13$, which still requires a full order of
magnitude variation of $k$. When $\beta\neq 0$, the only difference is
that the deficit window ends, roughly, when $X < X_\beta$.

The most significant parametrization for the \textbf{atan} model is the
special case \textbf{atan1}; Table~\ref{tab:atan1} summarizes our
findings for this model. We observe that adding
polarization data has an effect opposite to what we found for
\textbf{bpl}, with the value of $k^{-1}_\mathrm{c}$ getting larger as
we include more polarization datasets (independently of the other
datasets used, H0 and/or BAO). This occurs because the deficit actually
worsens the polarization fit, so the effect needs be reduced in order
to improve the temperature fit without spoiling the polarization fit.
Since this model has a single parameter, the only way to achieve that
is by pushing this parameter to a larger scale. Similarly to
\textbf{bpl}, when using temperature data only, the larger deficit is
compensated for by a 15\%-20\% increase in the spectrum amplitude.

Although this model seems to provide fits which are better than for the
fiducial case, in some cases with an improvement close to $2.5\sigma$,
these fits are actually not reliable because they are unstable with
respect to the parameter values obtained. Indeed, when polarization
data are added, the parameters change abruptly and, moreover, the
significance is reduced, as expected for a statistical fluctuation.
Finally, including all datasets, we obtain a marginal $1.8 \sigma$
significance for a smooth and broad deficit at very large scales.

\begin{table*}
\begin{ruledtabular}
\begin{tabular}{ l | c | c | c | c | c | c | c | c | c | c | c | c | r }
data-set & $k_\mathrm{c}^{-1} [\mathrm{Gpc}]$ & $\lambda$ & $\beta$ & $\alpha$ & $h$ & $\ns$ & $\ln(10^{10}\As)$ & $\tau$ & $\Gamma$ & $\gamma$ \\
\hline 	Planck $TT$  &  $ 0.36 $  &     --     &  $ 0.78 $  &     --     &  $ 0.69 $  &  $ 0.98 $  &  $ 3.3 $  &  $ 0.17 $  &  $ 6.8 $  &  $  3.4 $\% ($ 2.1 \sigma$) $[2]$\\
\hline 	Planck $TT$+H0  &  $ 0.35 $  &     --     &  $ 0.76 $  &     --     &  $ 0.7 $  &  $ 0.99 $  &  $ 3.3 $  &  $ 0.2 $  &  $ 7.2 $  &  $  2.8 $\% ($ 2.2 \sigma$) $[2]$\\
\hline 	Planck $TT$+BAO  &  $ 0.36 $  &     --     &  $ 0.79 $  &     --     &  $ 0.68 $  &  $ 0.97 $  &  $ 3.3 $  &  $ 0.16 $  &  $ 6.8 $  &  $  3.4 $\% ($ 2.1 \sigma$) $[2]$\\
\hline 	Planck $TT$+H0+BAO  &  $ 0.35 $  &     --     &  $ 0.77 $  &     --     &  $ 0.69 $  &  $ 0.98 $  &  $ 3.3 $  &  $ 0.18 $  &  $ 6.7 $  &  $  3.5 $\% ($ 2.1 \sigma$) $[2]$\\
\hline 	Planck $TT$+lowP  &  $ 0.37 $  &     --     &  $ 0.81 $  &     --     &  $ 0.68 $  &  $ 0.97 $  &  $ 3.1 $  &  $ 0.1 $  &  $ 5.1 $  &  $  8 $\% ($ 1.8 \sigma$) $[2]$\\
\hline 	Planck $TT$+lowP+H0  &  $ 0.37 $  &     --     &  $ 0.81 $  &     --     &  $ 0.69 $  &  $ 0.97 $  &  $ 3.2 $  &  $ 0.12 $  &  $ 4.6 $  &  $  10 $\% ($ 1.6 \sigma$) $[2]$\\
\hline 	Planck $TT$+lowP+BAO  &  $ 0.37 $  &     --     &  $ 0.8 $  &     --     &  $ 0.68 $  &  $ 0.97 $  &  $ 3.2 $  &  $ 0.12 $  &  $ 4.3 $  &  $ 11 $\% ($ 1.6 \sigma$) $[2]$\\
\hline 	Planck $TT$+lowP+H0+BAO  &  $ 0.37 $  &     --     &  $ 0.81 $  &     --     &  $ 0.69 $  &  $ 0.97 $  &  $ 3.2 $  &  $ 0.12 $  &  $ 4.8 $  &  $  9.2 $\% ($ 1.7 \sigma$) $[2]$\\
\hline 	Planck $TT$, $TE$, $EE$+lowP  &  $ 0.38 $  &     --     &  $ 0.82 $  &     --     &  $ 0.67 $  &  $ 0.96 $  &  $ 3.1 $  &  $ 0.093 $  &  $ 5.8 $  &  $  5.5 $\% ($ 1.9 \sigma$) $[2]$\\
\hline 	Planck $TT$, $TE$, $EE$+lowP+H0  &  $ 0.39 $  &     --     &  $ 0.82 $  &     --     &  $ 0.68 $  &  $ 0.97 $  &  $ 3.1 $  &  $ 0.095 $  &  $ 5.3 $  &  $  7.1 $\% ($ 1.8 \sigma$) $[2]$\\
\hline 	Planck $TT$, $TE$, $EE$+lowP+BAO  &  $ 0.38 $  &     --     &  $ 0.83 $  &     --     &  $ 0.68 $  &  $ 0.97 $  &  $ 3.1 $  &  $ 0.093 $  &  $ 5.3 $  &  $  7.1 $\% ($ 1.8 \sigma$) $[2]$\\
\hline 	Planck $TT$, $TE$, $EE$+lowP+H0+BAO  &  $ 0.38 $  &     --     &  $ 0.83 $  &     --     &  $ 0.68 $  &  $ 0.97 $  &  $ 3.1 $  &  $ 0.095 $  &  $ 5.1 $  &  $  7.7 $\% ($ 1.8 \sigma$) $[2]$\\
\end{tabular}
\end{ruledtabular}
\caption{\label{tab:jump} Excerpt from Tables~\ref{tab:TT_onlyMP} to \ref{tab:TTTEEE_lowP_H0_BAOMP} selecting the results for \textbf{jump}, see Table~\ref{tab:TT_onlyMP} for more details.}
\end{table*}

\subsection{Sharp deficit model: \textbf{jump}}

The results for this model are summarized in Table~\ref{tab:jump}. We
observe that the parameter values are now consistent across all
datasets, although the significance remains marginal and resembles those
for \textbf{atan1} and \textbf{expc2}. In the case of \textbf{jump},
however, all datasets point in the same general direction and parameter
values that provide a good fit for one dataset also provide a good fit
for the others. This means that, to have a hope of measuring an effect
using the full range of datasets, we need to consider the \textbf{jump}
model.

\bibliographystyle{apsrev}
\bibliography{references}

\end{document}